\documentclass[acmtog,preprint]{acmart}

\hypersetup{draft}

\usepackage{booktabs} % For formal tables

\usepackage[ruled]{algorithm2e} % For algorithms

\usepackage{todonotes}
\usepackage{multirow}
\usepackage{amsmath}

\SetAlFnt{\small}
\SetAlCapFnt{\small}
\SetAlCapNameFnt{\small}
\SetAlCapHSkip{0pt}

\setcopyright{acmlicensed}
\acmJournal{TOG}
\acmSubmissionID{394}
\begin{document}

\title{Finding Hexahedrizations for Small Quadrangulations of the Sphere}

\author{Kilian Verhetsel}
\orcid{0000-0001-7377-3491}
\affiliation{%
  \institution{Universit\'e catholique de Louvain}
  \streetaddress{Avenue Georges Lema\^itre 4-6}
  \city{Louvain-la-Neuve}
  \postcode{1348}
  \country{Belgique}
}%
\email{kilian.verhetsel@uclouvain.be}

\author{Jeanne Pellerin}
\orcid{0000-0001-8481-7509}
\affiliation{%
  \institution{Universit\'e catholique de Louvain}
  \streetaddress{Avenue Georges Lema\^itre 4-6}
  \city{Louvain-la-Neuve}
  \postcode{1348}
  \country{Belgique}
}%
\email{jeanne.pellerin@total.com}

\author{Jean-Fran\c cois Remacle}
\orcid{0000-0002-4798-6458}
\affiliation{%
  \institution{Universit\'e catholique de Louvain}
  \streetaddress{Avenue Georges Lema\^itre 4-6}
  \city{Louvain-la-Neuve}
  \postcode{1348}
  \country{Belgique}
}%
\email{jean-francois.remacle@uclouvain.be}

\begin{abstract}
  This paper tackles the challenging problem of constrained hexahedral
  meshing. An algorithm is introduced to build combinatorial hexahedral meshes
  whose boundary facets exactly match a given quadrangulation of the topological
  sphere.  This algorithm is the first practical solution to the problem.  It is
  able to compute small hexahedral meshes of quadrangulations for which the
  previously known best solutions could only be built by hand or contained
  thousands of hexahedra.  These challenging quadrangulations include the
  boundaries of transition templates that are critical for the success of
  general hexahedral meshing algorithms.
   
  The algorithm proposed in this paper is dedicated to building combinatorial
  hexahedral meshes of small quadrangulations and ignores the geometrical
  problem. The key idea of the method is to exploit the equivalence between quad
  flips in the boundary and the insertion of hexahedra glued to this
  boundary. The tree of all sequences of flipping operations is explored,
  searching for a path that transforms the input quadrangulation $Q$ into a new
  quadrangulation for which a hexahedral mesh is known. When a small hexahedral
  mesh exists, a sequence transforming $Q$ into the boundary of a cube is found;
  otherwise, a set of pre-computed hexahedral meshes is used.

  A novel approach to deal with the large number of problem symmetries is
  proposed. Combined with an efficient backtracking search, it allows small
  shellable hexahedral meshes to be found for all even quadrangulations with up
  to $20$ quadrangles. All $54,943$ such quadrangulations were meshed using no
  more than $72$ hexahedra. This algorithm is also used to find a construction
  to fill arbitrary domains, thereby proving that any ball-shaped domain bounded
  by $n$ quadrangles can be meshed with no more than $78~n$ hexahedra. This
  very significantly lowers the previous upper bound of $5396~n$.
\end{abstract}

%
% The code below should be generated by the tool at
% http://dl.acm.org/ccs.cfm
%



%
% End generated code
%


\begin{teaserfigure}
  \centering
  \includegraphics[width=0.95\linewidth]{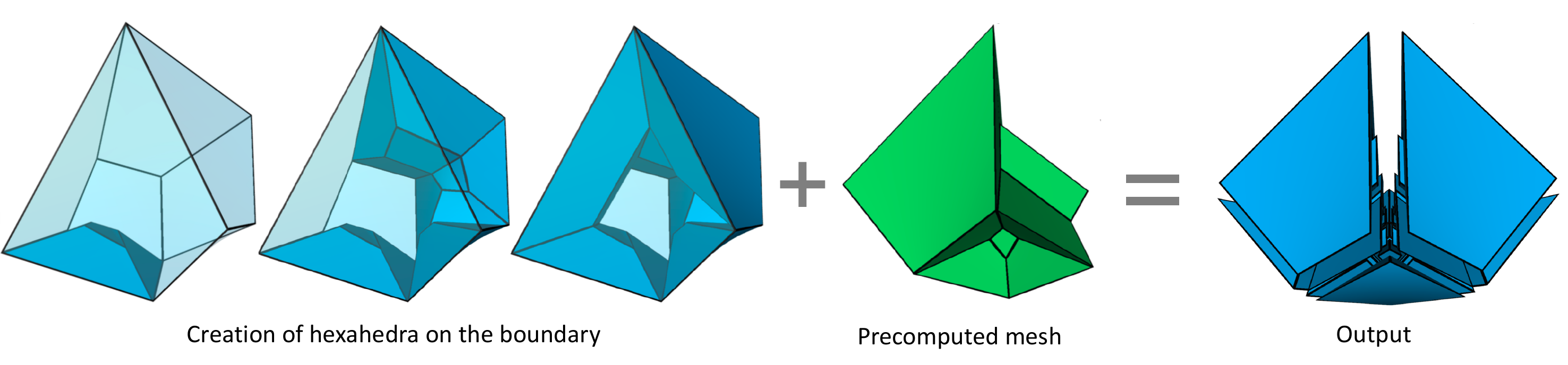}~%
  \caption{Given a quadrangulation of the topological sphere, our algorithm creates
    hexahedra on the boundary until the unmeshed cavity 
    matches the boundary of a pre-computed hex mesh that is merged 
    to obtain the final combinatorial hexahedral mesh.}
  \label{fig:teaser}
\end{teaserfigure}

\maketitle

\section{Introduction}

Volumetric mesh generation is a required step for engineering analysis. Robust
algorithms are able to automatically produce a tetrahedral mesh constrained to
have a given triangulation as its boundary, e.g. \cite{si2015tetgen}.  However,
subdivisions into hexahedra (cube-like cells) are often preferred over
tetrahedrizations for their good numerical properties such as a better convergence with fewer
elements \cite{Shepherd-2008} and faster assembly times \cite{Remacle-2016}. Yet,
the hexahedral meshing problem, and more particularly the boundary constrained
variant, remains open to this date.

Finding solutions to the boundary constrained hex-meshing problem is crucial for
hex-meshing algorithms that use a few simple templates to reduce the complexity of the
general meshing problem to a small set of inputs (\emph{e.g.}
\cite{mitchell1999geode, Yamakawa-2002}). More importantly, hex-dominant mesh
generation techniques usually leave small cavities unmeshed \cite{Yamakawa-2003}
and filling them is one of the missing pieces to the more
general problem of all-hex mesh generation.

This paper introduces an algorithm that solves the combinatorial constrained
hex-meshing problem for small quadrangulation of the sphere
(\autoref{fig:teaser}). Given a quadrangulation of the sphere $Q$, it determines a set of
combinatorial cubes $H$ such that:

\begin{enumerate}
 \item the intersection of any two hexahedra $h_1, h_2 \in H$ is 
  a combinatorial face shared by $h_1$ and $h_2$ (\emph{i.e.} the empty set, a
  vertex, an edge, or a quadrangle);
\item all quadrangular faces are shared by at most two hexahedra; and
\item the set of boundary quadrangle faces (adjacent to exactly one hexahedron) is
  equal to $Q$.
\end{enumerate}

\begin{figure}
  \centering
  \includegraphics[width=0.5\linewidth]{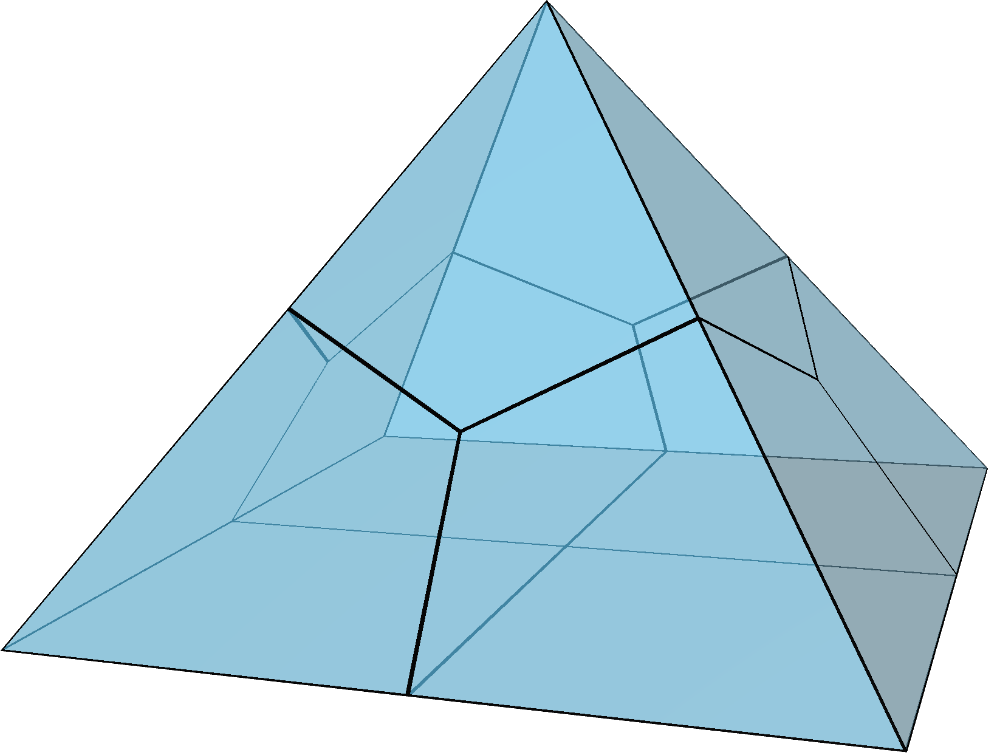}
  \hfill{}~%
  \hspace{0.08\linewidth}
  \includegraphics[width=0.3\linewidth]{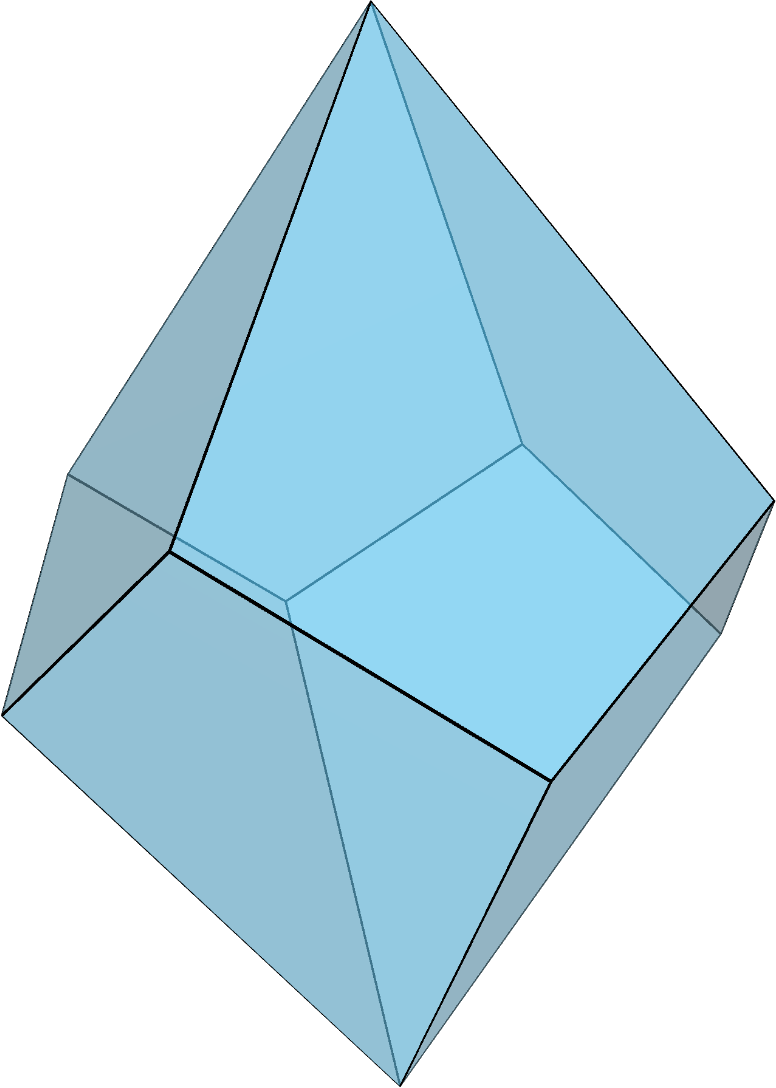}
  \caption{Schneiders' pyramid \cite{Schneiders-1996} and the tetragonal
    trapezohedron are two extremely challenging boundaries for combinatorial
    constrained hexahedral mesh generation.}
  \label{fig:pyramid-spindle}
\end{figure}

This is an extremely challenging problem, even when the subsequent problem of
finding a geometrical embedding is ignored, and for which no practical method
exists.
The existence of hexahedral meshes for all even quadrangulation of the
topological sphere has been proven by \citet{Mitchell-1996}, yet seemingly
innocuous quadrangulations such as the 16-quadrangle pyramid (Schneider's
pyramid) or the 8-quadrangle tetragonal trapezohedron
(\autoref{fig:pyramid-spindle}) are notorious failure cases of general purpose
meshing methods.

\citet{Eppstein-1999} shows how the interior of a quadrangulated sphere can be
meshed with a linear number of hexahedra; this construction was later
generalized to all domains which admit hexahedral meshes
\cite{Erickson-2014}. Both methods reduce the problem to meshing a few
quadrangulated spheres, but neither provide explicit hex meshes for these
cases.  Previous attempts to mesh these templates have been unsuccessful
\cite{Mitchell-Courses, Weill-Tetrahedron}.

One method only is able to generate hexahedral meshes for the templates of
Eppstein and Erickson \cite{Carbonera-2010}. The drawback is that it requires
$5396~n$ hexahedra to construct a non-degenerate hexahedral mesh of a ball
bounded by $n$ quadrangles.
%, and is not directly applicable to arbitrary domains. %accessoire. le gros chiffre suffit 
% pour démontrer que notre contribution tue tout

\paragraph{Contributions}

The first contribution of this paper is a practical algorithm to build
combinatorial hexahedral meshes of reasonable size for small quadrangulation of
the sphere. The algorithm is based on \emph{quad flips}, a set of operations to
modify quadrilateral meshes and whose application can be interpreted as the
construction of a hexahedron. Given a quadrangulated sphere $Q$, a hexahedral
mesh bounded by $Q$ is built by exploring the space of flipping operations that
can be applied to $Q$. A solution is then obtained by finding a sequence of
operations that transforms $Q$ into the boundary of a cube
(\autoref{sec:search}). When this search space is too large, the algorithm
instead searches for a sequence of operations transforming $Q$ into the boundary
of any mesh within a library of pre-computed hexahedral meshes
(\autoref{sec:look-ahead}).

This algorithm is used to construct combinatorial hexahedral meshes for all
$54,943$ quadrangulations of the sphere with up to $20$ quadrangles and which
admit a hexahedral mesh. The computed hexahedral meshes contain at most $72$
hexahedra.

The last contribution of this work is to significantly lower the upper bound
needed to mesh arbitrary domains. The construction of Erickson is made fully
explicit by computing hexahedral meshes for its two quadrangulated templates. This
proves that an arbitrary ball bounded by $n$ quadrangles can be meshed using
only $78~n$ hexahedra.

An implementation of all algorithms introduced in this paper is provided as free
software and can be found in the supplementary materials or from
\url{https://www.hextreme.eu}.

\section{Related Work}  

Hexahedral mesh generation is a thriving field of research, with a variety of
proposed methods. These include multi-block decomposition methods using
frame-field parametrizations \cite{Kowalski-2014, Nieser-2011, Liu-2018,
  Lyon-2016}, hex-dominant meshing methods \cite{Yamakawa-2003, Gao-2017,
  Baudouin-2014, Sokolov-2017, Pellerin-2018}, octree-based methods \cite{Marechal-2009,
  Ito-2009, Zhang-2012, Qian-2010}, and polycube-based methods
\cite{Gregson-2001, Han-2011, Yu-2014, Fang-2016}.

A detailed overview of these various approaches is beyond the scope of this
paper, as most methods do not address the problem of generating meshes with a
given boundary quadrangular mesh. The rest of this section focuses on methods
tackling the constrained problem.

\subsection{Existence proofs}

\begin{figure}
  \centering
  \includegraphics[width=0.6\linewidth]{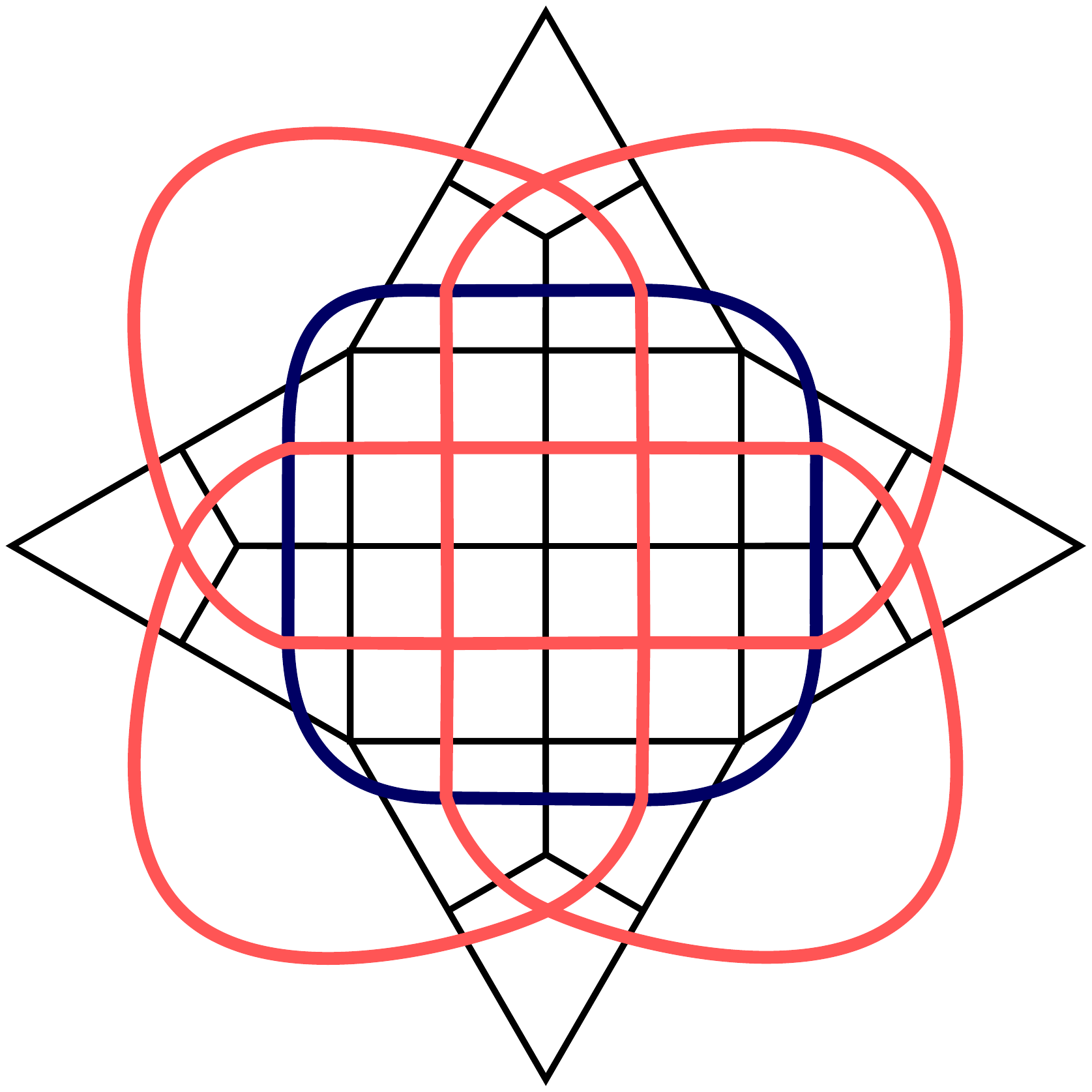}
  \caption{The dual of a quadrangulated mesh (black edges) can be seen as an arrangement of two
    curves. The dark blue curve is simple, the red curve intersects itself 8 times. 
    This example mesh is Schneider's pyramid.}
  \label{fig:dual-pyramid}
\end{figure}

\paragraph{Existence theorem for ball inputs} \citet{Thurston-1993} and
\citet{Mitchell-1996} independently showed that a ball bounded by a
quadrangulated sphere can be meshed with hexahedra if and only if the number of
quadrangles on the boundary, $n$, is even. The proof is based on the dual cell
complex of quadrangular and hexahedral meshes.  The dual complex of a
quadrangular mesh is obtained by placing a vertex at the center of each
quadrangle and adding edges between the vertices corresponding to adjacent
quadrangles.  Grouping edges traversing opposite edges of a same quadrangle, the
dual complex is interpreted as an arrangement of curves
(\autoref{fig:dual-pyramid}).  Similarly, the dual of a hexahedral mesh can be
interpreted as an arrangement of surfaces \cite{Murdoch-1997}.  In Mitchell's
proof, an arrangement of surfaces bounded by the dual arrangement of curves of
the input quadrangulation is first constructed. For curves with an even number
of self-intersections (including curves with no self-intersections), a disk is
constructed inside the domain and a regular homotopy between a circle and the
curve can be used to create a manifold bounded by that curve. Curves with an odd
number of self-intersections are paired up arbitrarily. For each pair, a
manifold bounded by the two curves is constructed by computing a regular
homotopy between the two of them. This arrangement is not in general the dual of
a hexahedral mesh, so the next step of the construction is to add new surfaces
completely inside the ball until all connectivity requirements of a hexahedral
mesh are met.

\paragraph{Linear-complexity meshing} The construction of Mitchell can
necessitate up to $\Omega(n^2)$ hexahedra where $n$ is the number
quadrangles. Eppstein showed this was the case and proposed a different
construction which guarantees the use of $O(n)$ hexahedra \cite{Eppstein-1999}.
Eppstein's algorithm first subdivides each quadrangle into two triangles, so
that a tetrahedral mesh of the interior can be computed. After subdividing each
tetrahedron into four hexahedra, a hexahedral mesh is obtained.  However, its
boundary does not match the initial input quadrangulation. This is solved by
inserting \emph{buffer cells}: for each quadrangle, add a cube, and glue one of
its face to the original quadrangle; then, subdivide the opposite face into six
quadrangles. The six new quadrangles are matched with those obtained from
subdividing the original quadrangles during the previous step. The four
remaining sides of the buffer cells are carefully subdivided into either two or
three quadrangles, so that each buffer cell is bounded by an even number of
quadrangles. Mitchell's proof can then be invoked to show that each buffer cell
can be subdivided into a finite number of hexahedra.

\begin{figure}
  \includegraphics[width=0.5\linewidth]{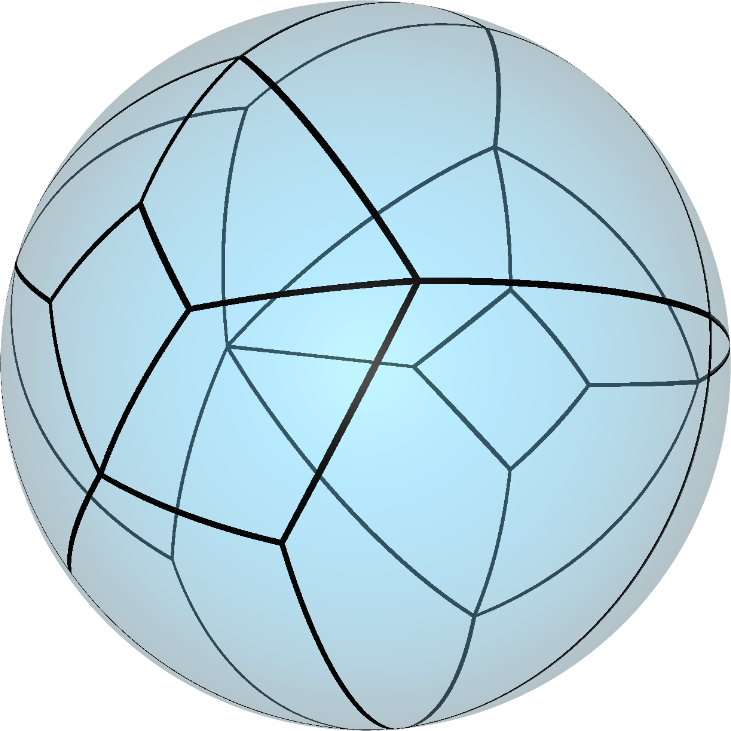}~
  \includegraphics[width=0.5\linewidth]{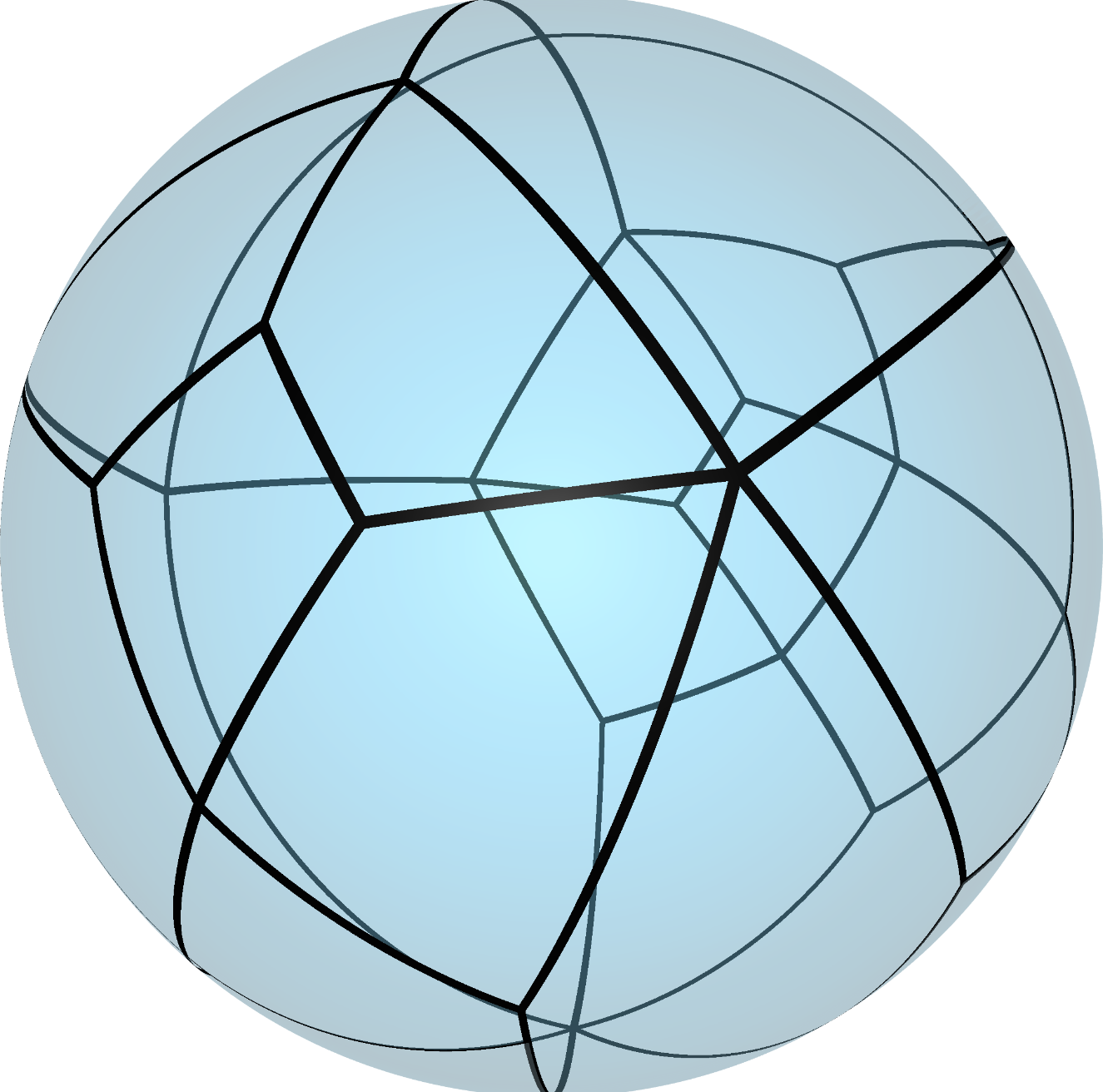}
  \caption{If hexahedrizations of these two quadrangulated spheres exists,
    then it is possible to construct
    hexahedrizations for all other quadrangulated surface \cite{Erickson-2014}.}
  \label{fig:erickson-buffers}
\end{figure}

\paragraph{Generalization to other inputs} Generalizing the previous results,
\citet{Erickson-2014} gives necessary and sufficient conditions for the
existence of a hexahedral mesh of a domain $\Omega$ bounded by a quadrangulation
$Q$.  The requirement is that every null-homologous subgraph of the input
quadrangulation (\emph{i.e.} every subgraph which bounds an embedded surface of
$\Omega$) contain an even number of edges. The construction of Erickson is
similar to the one proposed by Eppstein, and also starts by computing a
tetrahedral mesh of the domain, subdividing it into a hexahedral mesh, and
inserting buffer cells to get a complete mesh with the correct boundary. The
last step is to subdivide the buffer cells of two different types
(\autoref{fig:erickson-buffers}) into hexahedra, which is again shown to be
possible from Mitchell's proof.  Like Eppstein, Erickson does not give an
explicit construction of the hexahedral meshes of the buffer cells that are the
base of its proofs.

\paragraph{A constructive method}
\citet{Carbonera-2010} give the first completely explicit construction. Their
algorithm first adds hexahedra inside the domain, guaranteeing that the dual
arrangement of the boundary of the remaining region contain no self-intersecting
curve. Buffer cells are then inserted to transition to a mesh where each
quadrangle has been subdivided into four quadrangles. The rest of the domain is
then filled using pyramids. A complete hexahedral mesh is obtained after
subdividing the pyramids into hexahedra. Given a topological ball bounded by $n$
quadrangles, their construction produces a mesh of $76~n$ hexahedra. This mesh
is degenerate: it contains quadrangles sharing multiple edges and hexahedra
sharing multiple faces. A combinatorially valid mesh can be obtained by further
refining the mesh \cite{Mitchell-1995}. This method, however, requires as many
as $5396~n$ hexahedra to build a hexhahedral mesh bounded by quadrangulation of
size $n$.

\subsection{Constrained hexahedral meshing in practice}

The methods in the previous section are impractical, they create far more
hexahedra than necessary. In practice, less general methods have been used to
obtain smaller meshes.

\paragraph{Whisker Weaving} Whisker Weaving has been proposed by
\cite{Tautges-1996}.  The idea is to use a topological advancing front to
construct the dual of a hexahedral mesh. The algorithm initially assumes that
the final mesh will contain one dual surface for each dual curve of the input
quadrangulation. Hexahedra are created inside the domain by creating
intersections between three of these sheets, until the entire domain is
filled. To choose between the multiple possible operations, heuristics based on
geometric information such as the dihedral angle of faces are used. These
heuristics are often not enough to completely fill the domain.

\paragraph{Dual cycle elimination} \citet{Muller-1999} proposed a method based
on \emph{dual cycle eliminations}. At each step of the algorithm, one of the
curves of the dual mesh is removed, matching this elimination as the insertion
of a layer of hexahedra. The new boundary after removing this cycle bounds the
part of the input domain which has not been meshed yet. This process is repeated
until the boundary matches that of a single cube. This method succeeds for
certain classes of input quadrangulations, but fails for the common cases where
the dual contains self-intersecting curves (e.g. \autoref{fig:dual-pyramid}).

\subsection{Searching for hexahedral meshes}

Some specific cases have particularly attracted the attention of the research
community: Schneiders' pyramid \cite{Schneiders-Open} and the octogonal spindle
(\autoref{fig:pyramid-spindle}). The \textit{ad-hoc} constructions for the
pyramid proposed \citet{Yamakawa-2002, Yamakawa-2010} have long been the smallest
known solutions.  Recently, computer searches have been used to
find substantially smaller solutions.

\citet{Verhetsel-2018} exhaustively explore the space of all possible hexahedral
meshes up to a given number of vertices by considering the possible groups of 8
vertices that can be built without creating an invalid mesh.  The search space
is usually too large for a solution to be found except in fairly simple
cases. Starting from the 88-element solution of \citeauthor{Yamakawa-2010} and
successively coarsening small parts of the original mesh, the method of
\citeauthor{Verhetsel-2018} allowed the construction of a 44-element hex mesh of
the pyramid.

By considering a smaller search space, \citet{Xiang-2018} construct a mesh of
the pyramid with 36 hexahedra by building a shelling of the resulting hexahedral
mesh. Starting from a single cube, the algorithm of \citeauthor{Xiang-2018}
considers all possible ways to add one hexahedron while maintaining a mesh which
is both valid and combinatorially equivalent to a topological ball. The process
is stopped when the boundary of the mesh matches the target quadrangulation.

\section{Exhaustive Search}
\label{sec:search}

\subsection{Overview}

Given a quadrangulation of the sphere $Q$, we describe an algorithm to enumerate
all hexahedral meshes bounded by $Q$ and which can be constructed using quad
flips (\autoref{alg:search-exhaustive}). To force the algorithm to terminate,
the search is limited to meshes with a maximum of $H_{\max}$ hexahedra and a
maximum $V_{\max}$ of vertices. In \autoref{sec:look-ahead}, we then extend the
approach to search for hexahedral meshes that are prohibitively large for an
exhaustive search of this kind.

\begin{algorithm}[t]
  \caption{$\textsc{Search}$: Enumerate shellable hexahedral meshes}
  \label{alg:search-exhaustive}
  \KwIn{$Q$: A quadrangulation of the sphere; \\
        $H$: A partial mesh; \\
        $H_{\max}$: maximum number of hexahedra in a solution; \\
        $V_{\max}$: maximum number of vertices in a solution.}
  \lIf{$H_{\max} = |H|$}{\Return}
  \uElseIf{$\textsc{Visited-Symmetric-Counterpart}(H)$}{
    \Return \tcp*[r]{\autoref{sec:symmetry}}
  }
  \ElseIf{$Q \approx$ a cube}{
    $h \gets $ the hexahedron bounded by $Q$ \tcp*[r]{\autoref{sec:flip}}
    \lIf{$\textsc{Is-Compatible}(H, h)$}{
    $\textsc{Output-Solution}(H \cup \{h\})$}
  }
  \ForEach{quad flip $F$}{
    $(Q', h) \gets \textsc{Perform-Flip}\left(F, Q\right)$ \tcp*{\autoref{sec:flip}}
    \lIf{$\textsc{Num-Vertices}(H \cup \{h\}) \ge V_{\max}$}{\textbf{continue}}
    \If{$\textsc{Is-Compatible}(H, h)$}{
      $\textsc{Search}(Q', H \cup \{h\}, H_{\max}, V_{\max})$\;
    }
  }
\end{algorithm}

The algorithm detailed in this section does not search the entire space of
hexahedral meshes, but only the space of so-called \emph{shellable} meshes,
which can be explored efficiently using quad flips (\autoref{sec:shelling}).
This space is explored in its entirety by considering all possible sequences of
quad flips that correspond to valid hexahedral meshes
(\autoref{sec:flip}). Because many different sequences of flipping operations
represent the same mesh, most of this section focuses on how to account for the
symmetries of the input quadrangulation, in order to avoid generating different
sequences of quad flips corresponding to isomorphic hexahedral meshes
(\autoref{sec:symmetry}).

\subsection{Shellability and quad flips}
\label{sec:shelling}

\begin{figure}
  \centering
  \includegraphics[width=\linewidth]{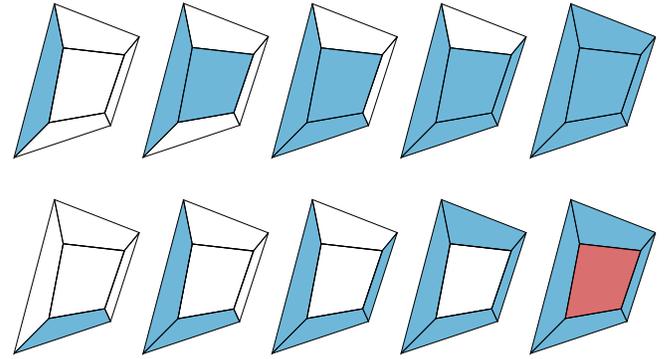}
  \caption{(top) a shelling of a quadrangulation; (bottom) not a shelling
    because a hole is present after inserting the first four quadrangles.}
  \label{fig:shelling}
\end{figure}

Our method only considers a specific class of meshes: \emph{shellable hexahedral
  meshes}. \emph{Shellability} is an important and useful combinatorial concept
in the study of polytopes and cell complexes \cite{Ziegler-1995}. Slightly
different notions of shellability are found in the literature. We use that of
\emph{pseudo-shellings} \cite{Bern-2002} or \emph{topology-preserving shellings}
\cite{Muller-1999}. This type of shelling is an ordering of the hexahedra
$(H_1, H_2, \dots, H_n)$ of a hexahedral mesh such that any prefix
$\bigcup_{0 \le i < k} H_i$ is homeomorphic to a ball (\autoref{fig:shelling}).

\begin{figure}
  \includegraphics[width=\linewidth]{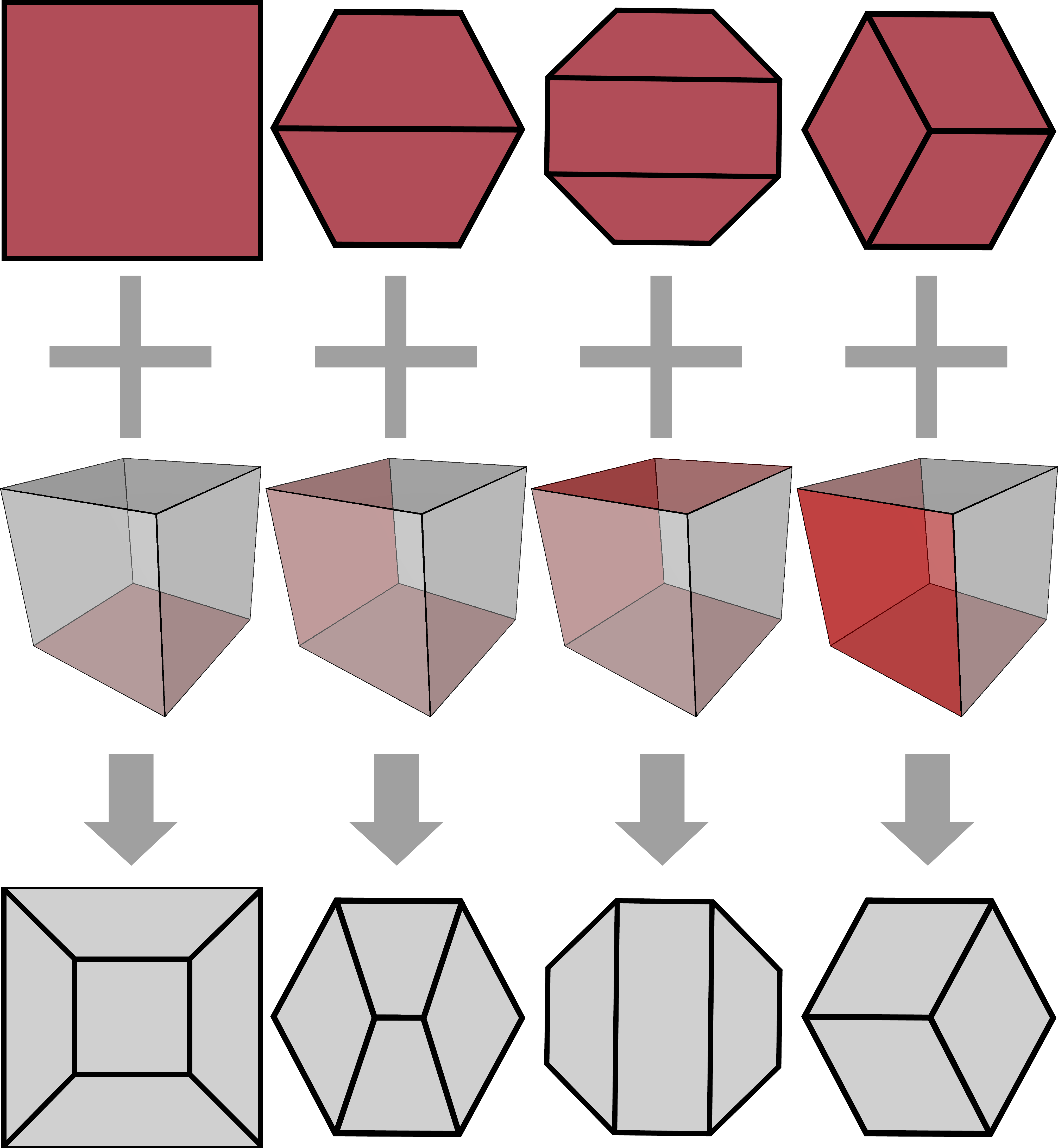}
  \caption{Equivalence between quad flips and hex creation. 
    Adding one hexahedron glued to the top row red quadrangles, 
    modifies locally the quads and results in the bottom row grey quadrangles.
    This operation is the key idea of the algorithm to search shellable meshes.}
  \label{fig:quad-flips}
\end{figure}

This definition implies that any hexahedron $H_k$ must intersect the union of
the previous hexahedra in one of six possible patterns. Gluing a hexahedron to
one of these patterns modifies the boundary of the mesh locally
(\autoref{fig:quad-flips}). The transitions between these patterns are known as
\emph{quad flips} or \emph{bubble moves} \cite{Funar-1999}. These flipping
operations are therefore a valuable building block to explore the space of
shellable meshes.

Note that not all hexahedral meshes admit a shelling order --- see for example
Furch's ball \cite{Furch-1924}. Hence, by relying on these flipping operations
to build hexahedral meshes, our method is inherently unable to construct certain
meshes. Nonetheless, we can guarantee that a solution still exists: all
quadrangulations of the sphere with an even number of quadrangles admit a
shellable hexahedral mesh \cite{Bern-2002}.

\subsection{Identifying and Performing Flips}
\label{sec:flip}

For each quadrangulation $Q$ visited during the search, all possible quad flips
need to be identified. Each flip corresponds to a different hexahedron that
can be inserted in the mesh. The algorithm successively tries adding all of them
to the current mesh. Because flips are performed by starting from the target
boundary, the hexahedra that are constructed during this process form the
reverse of a shelling order (this is one difference with the search method of
\citet{Xiang-2018}). \citet{Muller-1999} construct the hexahedra in 
the same order, 
but our method, instead of only considering one mesh, explores the
entire tree of possible sequences of quad flips.

The identification of all possible flips is split into two steps: first, the
boundary $Q$ is inspected to identify all occurrences of the 6 patterns from
\autoref{fig:quad-flips}. Second, those flips that correspond to the insertion
of hexahedra that would make the mesh invalid are filtered out.

The hexahedron inserted by performing a flip is obtained by computing the union
of the pattern before and after the flip. To determine whether or not this
hexahedron is compatible with the mesh constructed so far, an efficient test is
devised by considering three relations between the vertices of the mesh:
\begin{enumerate}
\item $E$, the edges of the mesh;
\item $D_{Q}$, the diagonals of the quadrangles in the mesh;
\item $D_{H}$, the interior diagonals of the hexahedra in the mesh.
\end{enumerate}

These relations are disjoint in any combinatorial hexahedral mesh. For example,
if a pair $(u, v)$ is an edge, it is not the diagonal of any quadrangles or
hexahedra. This leads to an efficient implementation of the test: simply
maintain the three sets $E$, $D_{Q}$, and $D_{H}$, and verify that, after adding
a new hexahedron:
\begin{enumerate}
\item the three sets $E$, $D_{Q}$, and $D_{H}$ remain disjoint; \label{item:disjoint}
\item the new quadrangles in the hexahedron share no diagonals with any
  other quadrangle in the mesh; \label{item:quad-diagonals}
\item none of the four interior diagonals of the new hexahedron are an interior
  diagonal of some other hexahedron. \label{item:hex-diagonals}
\end{enumerate}

It is easy to verify that when any two hexahedra share only a vertex, an edge,
or a quadrangle, these conditions are met. To verify their sufficiency, consider
two hexahedra with an invalid intersection pattern. If an interior diagonal of
one hexahedron is contained in the other hexahedron, one of the rules is always
violated: rule \ref{item:hex-diagonals} is violated if it is also an interior
diagonal of the second hexahedron, and rule \ref{item:disjoint} is violated if
it is an edge or the diagonal of a quadrangle. The only remaining cases to
consider are those where the shared vertices are part of two distinct
quadrangles. In all of those cases, the diagonal of one of those quadrangles
appears in the other one. If it appears as an edge, rule \ref{item:disjoint} is
violated; if not, both quadrangles have a shared diagonal, violating rule
\ref{item:quad-diagonals}.

The insertion of the last hexahedron requires special treatment. This step does
not correspond to a quad flip: when the boundary of the unmeshed region is
isomorphic to the boundary of a cube, a hexahedron is inserted to finish the
mesh. Detecting whether or not the current boundary corresponds to that of the
cube is straightforward: simply verify that the boundary has exactly 6 faces.

\subsection{Symmetry}
\label{sec:symmetry}

There are many distinct sequences of quad flips which represent identical
hexahedral meshes. It is thus important to only consider a single representation
for each hexahedral mesh constructed during the search, lest most of the
computation time be spent generating different representations of equivalent
solutions.

One technique commonly used to deal with this type of issue is to define a
canonical representation for objects under constructions, so that all those that
belong to a given isomorphism class are transformed into the same representative
element \cite{Burton-2011, Brinkmann-2007}. A significant portion of the
execution time is then spent computing the canonical representations of partial
solutions, which may completely change after every operation
\cite{Jordan-2018}. The symmetry breaking method used within our algorithm
instead compares partial solutions directly, and exploits the tree-shaped
structure of the search in order to reuse results from previous computations.

The strategy described in this section is based on \emph{Symmetry Breaking via
  Dominance Detection} (SBDD) \cite{Fahle-2001}. Consider the search tree
explored by the algorithm: its nodes are partial meshes constructed during the
search, and edges correspond to the insertion of new hexahedra through quad
flips. The objective is to prune from this search tree nodes that correspond to
meshes that have already been explored (up to symmetry). This is accomplished
using the following steps:

\begin{enumerate}
\item first, the automorphism group of the input quadrangulation is pre-computed;
\item then, as the search tree is traversed, fully explored subtrees are encoded
  into a sequence $S$;
\item for each new node, we determine whether or not it should be pruned by
  comparing it against the nodes stored in $S$.
\end{enumerate}

\subsubsection{Computing the automorphism group}
\label{sec:automorphism-group}

\begin{algorithm}[t]
  \caption{$\textsc{Compute-Symmetry}$: Computes one symmetry from an initial assumption}
  \label{alg:symmetry}
  \KwIn{$Q$: A quadrangulation of the sphere; a quadrangle $q_0 \in Q$; $(x, y, z, w)$, the
    image of $q_0$}
  \KwOut{The symmetry $\sigma$ that maps $q_0$ to $(x, y, z, w)$}
  Initialize $\sigma$, mapping $q_0$ to $(x, y, z, w)$\;
  Initialize a queue with the 4 edges of $q_0$\;
  $\textsc{Visited}_A \gets \{q_0\}$\;
  $\textsc{Visited}_B \gets \{(x, y, z, w)\}$\;
  $\textsc{Seen} \gets \{q_0\}$\;
  \While{the queue is not empty}{
    Dequeue an edge $(a, b)$\;
    $q \gets$ the quadrangle containing $(a, b)$ and not in $\textsc{Visited}_A$\;
    $q' \gets$ the quadrangle containing $\sigma(a, b)$ and not in $\textsc{Visited}_B$\;

    \ForEach{vertex $v$ in $q$}{
      $v' \gets $ the corresponding vertex in $q'$\;
      \uIf{$\sigma(v)$ is undefined and $\sigma^{-1}(v')$ is undefined}{
        $\sigma(v) \gets v'$ \tcp*{Extend the map $\sigma$}
        $\sigma^{-1}(v') \gets v$\;
      }
      \ElseIf{$\sigma(v) \ne v'$ or $\sigma^{-1}(v') \ne v$}{
        \textbf{fail} \tcp*{Stop upon contradiction}
      }
    }

    \ForEach{edge $e$ of $q$}{
      $o \gets $ the quadrangle on the other side of $e$\;
      \If{$o$ has not been seen before}{
        $\textsc{Seen} \gets \textsc{Seen} \cup \{o\}$\;
        Enqueue $e$\;
      }
    }

    $\textsc{Visited}_A \gets \textsc{Visited}_A \cup \{q\}$\;
    $\textsc{Visited}_B \gets \textsc{Visited}_B \cup \{q'\}$\;
  }
  \KwRet{$\sigma$}\;
\end{algorithm}

\begin{figure}
  \centering
  \includegraphics[width=\linewidth]{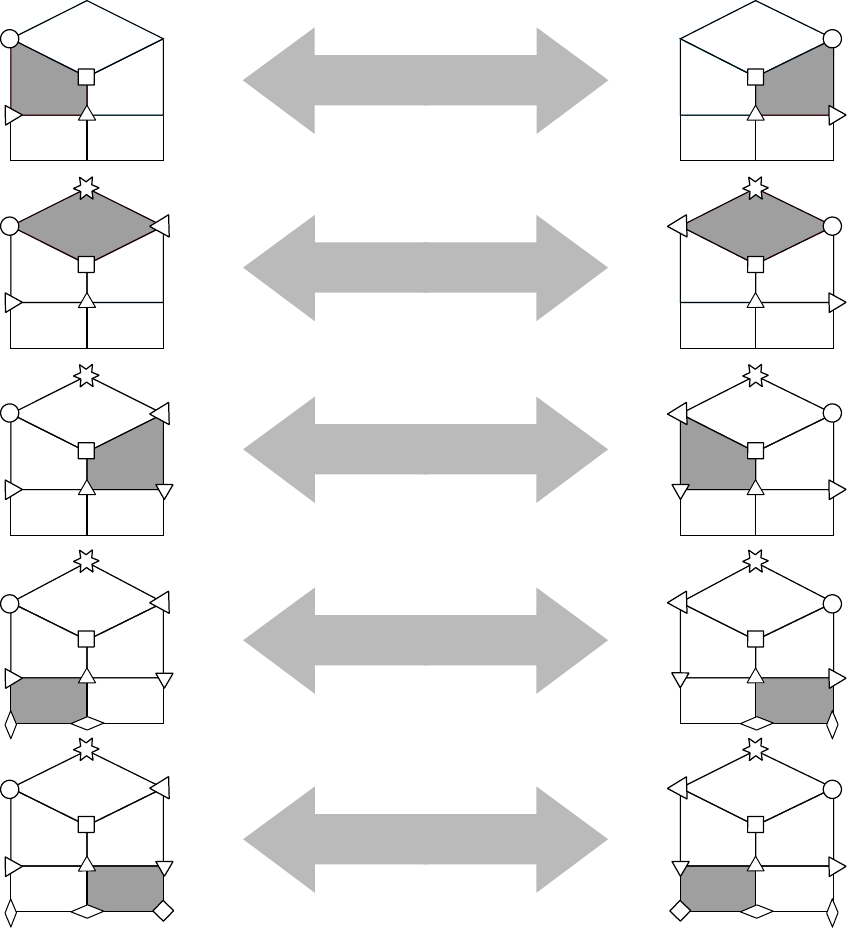}
  \caption{Computation of a symmetry. Starting from the assumption that a
    quadrangle is the image of some other quadrangle, the mesh is traversed
    while computing the correspondence between all other vertices.}
  \label{fig:symmetry-computation}
\end{figure}

Given a quadrangulation $Q$, we compute the set of its symmetries, known
as its automorphism group. A permutation $\sigma$ of the vertices of $Q$ is a
symmetry if it \emph{preserves} the set of quadrangles: for any quadrangle
$(a, b, c, d)$ of $Q$, its image $(\sigma(a), \sigma(b), \sigma(c), \sigma(d))$
is also quadrangle of $Q$, and every quadrangle $(a, b, c, d)$ is the image of a
quadrangle $(\sigma^{-1}(a), \sigma^{-1}(b), \sigma^{-1}(c), \sigma^{-1}(d))$.
Note that the orientations of the quadrangles may be reversed by $\sigma$.

Symmetries are computed one at a time, by fixing some quadrangle $q_A \in Q$ and
assuming that its image under a symmetry $\sigma$ is known to be $q_B \in
Q$. There are 8 different ways to map the vertices of $q_A$ onto the vertices of
$q_B$, corresponding to the 8 symmetries of a quadrangle. The entire permutation
$\sigma$ is uniquely determined by this part of the map
(\autoref{fig:symmetry-computation}): the quadrangles adjacent to $q_A$ must be
the images of the quadrangles adjacent to $q_B$ under $\sigma$, and the
quadrangles adjacent to those must also be images of each other, and so on,
until the whole quadrangulation has been traversed
(\autoref{alg:symmetry}). This process is well-defined because each edge is in
at most two quadrangles.

The entire set of symmetries is computed by considering all $8|Q|$ possible ways
to map an arbitrary quadrangle $q_A$ to any other quadrangle of $Q$. If an
assumption is correct, a symmetry $\sigma$ is obtained; if not, a contradiction
will be reached when trying to construct the symmetry (two vertices mapping onto
the same target vertex, or a single vertex with two images under
$\sigma$). Because $q_A$ must be the image of some quadrangle under any symmetry
$\sigma$, this process yields the entire automorphism group.

In the worst case, the entire automorphism group is determined in $O(|Q|^2)$
operations.  In practice, this quadratic time algorithm outperforms more complex
linear time algorithms designed for planar graph isomorphism
\cite{Eppstein-1999b, Colbourn-1981} when applied to small quadrangulations,
thanks to well-tuned heuristics. In particular, our implementation stops the
algorithm as soon as two vertices of different degree are mapped onto one
another by the permutation under construction \cite{Brinkmann-2007}.

Moreover, because this method does not use the planarity of the graph, it is
also more general. The only requirement is that the input be a
\emph{pseudomanifold}: a combinatorial cell complex in which every facet is
contained in at most two distinct cells. Indeed, a variant of this method will
be used to compare hexahedral meshes in section~\ref{sec:dominance}, by having
quadrangles take over the role of edges in \autoref{alg:symmetry}.

\subsubsection{Encoding the search tree}

\begin{figure}
  \centering
  \includegraphics[width=\linewidth]{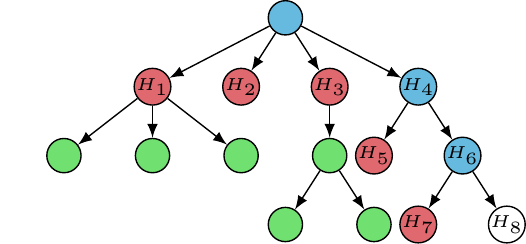}
  \includegraphics[width=\linewidth]{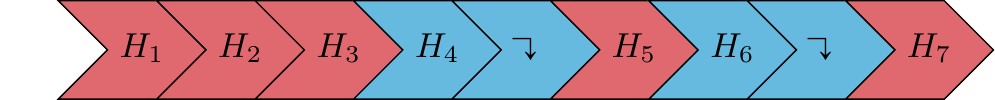}
  \caption{A partially explored search tree and the sequence used to compare the
    current node (in white) against the previously explored part of the
    tree. No-goods are shown in red, and partially explored subtrees in blue.}
  \label{fig:no-goods-sequence}
\end{figure}

An efficient traversal of the search tree requires the search to stop as soon as
the mesh under construction is the symmetric counterpart of a mesh that has
previously been constructed. In the previous section, the set of symmetries that
need to be considered was determined. This section now focuses on efficiently
encoding the set of hexahedral meshes that have been constructed during the
search.

Of course, the search tree is exponentially large, making it impossible to store
every single mesh that is constructed during the search. Instead, SBDD only
stores information about the roots of maximal fully explored subtrees, known as
\textit{no-goods} \cite{Gent-2006}. The current node should then be pruned if
and only if the mesh under construction is the symmetric counterpart of one of
the children of one of the no-goods. Note that a no-good is referred as such
even if some of its children are solutions, since it is not desirable to compute
the symmetric counterparts of those solutions.

No-goods can be stored efficiently thanks to the structure of the search tree.
Recall that each node within the search tree corresponds to a partial hexahedral
mesh, and each edge corresponds to the insertion of a hexahedron. Nodes with a common
ancestor in the tree then share a common set of hexahedra as a prefix, and this
prefix only needs to be stored once (\autoref{fig:no-goods-sequence}). Upon
visiting a new node, the most recently added hexahedron is inserted in the
sequence, followed by a special branching symbol, indicating that the rest of
the sequence will encode the children of this node. Upon backtracking,
everything up to and including the last branching symbol of the sequence is
removed.

\subsubsection{Dominance detection and pruning}
\label{sec:dominance}

The last part of our symmetry breaking method is the test used to prune nodes of
the search tree that do not need to be explored because any solution that could
be found by doing so has already been found. These nodes are said to be
\textit{dominated} by one of the no-goods, \textit{i.e.} they are the symmetric
counterpart of one of the children of one of the nodes that have been previously
explored and stored in the sequence shown in \autoref{fig:no-goods-sequence}.

Since the search involves exploring exponentially many nodes, this dominance
test must be implemented without explicitly comparing the current node against all
previously explored nodes. Instead, this test is broken down into two steps:
first find a no-good such that all its hexahedra are contained in the current partial
solution, then determine if the hexahedra that are in the partial
solution but missing from the no-good could be inserted using flipping
operations. The sequence $S$ constructed in the previous section is very
valuable for this: not only does it save space by factoring out a common prefix,
but it also saves time by allowing this prefix to be processed only once.

Let $H$ be the current partial solution. The first step is to search within $S$
for a partial mesh whose hexahedra are a subset of $H$
(\autoref{alg:dominance-test}). The process to find such a partial mesh is
similar to the algorithm used to compute the automorphism group initially
(section~\ref{sec:automorphism-group}). The goal is to construct $\sigma$, which
maps the vertices of some partial mesh encoded in $S$ to vertices of the current
solution $H$, such that all hexahedra in the no-good are preserved by the map
$\sigma$. The construction of $\sigma$ again begins from an initial assumption,
namely that the images of all boundary vertices through $\sigma$ are
known. Because boundary quadrangles must be preserved by $\sigma$, the set of
possible initial assumptions is precisely the automorphism group that was
previously computed.

Algorithm~\ref{alg:dominance-test} is executed once for each element of the
automorphism group and consists in a traversal of $S$ during which the map
$\sigma$ is extended. The process ends either upon finding a partial mesh
contained in $H$ or upon reaching a contradiction. For each hexahedron $h$ found
in $S$, we attempt to extend $\sigma$ such that $h$ maps to some hexahedron of
the current solution $H$. Each hexahedron created by a quad flip shares at least
one quadrangle with the boundary or with a previously created
hexahedron. Because of this, each hexahedron in $S$ has at least one quadrangle
whose symmetric counterpart is known. It is therefore possible to search for the
hexahedron $h'$ within the current solution $H$ that contains this quadrangle
(and has not already been determined to be the symmetric counterpart of another
hexahedron).

If such a hexahedron $h'$ exists, it must be the symmetric counterpart of the
hexahedron $h$ encoded in $S$, and the map $\sigma$ is extended accordingly. If
$h$ corresponds to a fully explored node (shown in red in
\autoref{fig:no-goods-sequence}), the current partial solution $H$ contains the
symmetric counterparts of all hexahedra of the corresponding no-good. If,
however, $h$ corresponds to a partially explored node (shown in blue in
\autoref{fig:no-goods-sequence}), and its symmetric counterpart cannot be found,
the traversal ends early because all subsequent partial meshes encoded in $S$
contain $h$, which is not in the current solution. In all other cases, the
traversal of the sequence continues.

Clearly, containing all hexahedra from some no-good is a requirement for a node
being dominated --- all children of the no-good share this common prefix. There
could still be cases where none of the children of this no-good contain all the
hexahedra that are in the current node. In other words, it may be impossible to
find a sequence of quad flips which inserts the missing hexahedra when starting
from the no-good. Testing for the existence of such a sequence may appear
intractable at first, because shellability is an NP-Complete property
\cite{Goaoc-2018}. Thankfully, a correct test only needs not to produce any
false positives, since false positives are the only reason a part of the search
tree would incorrectly get pruned, causing solutions to be missed. Furthermore,
because shellable meshes tend to accept many different shelling orders, there is
a straightforward algorithm meeting this requirement and which very often
computes the correct result: try a small number of permutations (say 10), then
give up if no reverse shelling order was found (\autoref{alg:reverse-shell}).

\begin{algorithm}
  \caption{$\textsc{Contains-No-Good}$: Determine whether or not a partial mesh
    contains the symmetric counterpart of a previously visited node}
  \label{alg:dominance-test}
  \KwIn{%
    $S$: an encoding of the part of the search tree explored so far
    (\autoref{fig:no-goods-sequence}); $H$: a partial mesh; $\sigma$: a symmetry
    of the target boundary.
  }
  \KwOut{%
    $\mathbf{true}$ if $H$ contains the symmetric counterparts of all the
    hexahedra of a fully explored subtree encoded in $S$
  }

  $\textsc{Seen} \gets \emptyset$\;
  \ForEach{hexahedron $h \in S$}{
    $\textsc{Success} \gets \mathbf{true}$\;

    $q \gets $ a quadrangle of $h$ whose image through $\sigma$ is known\;
    $q' \gets \sigma(q)$\;

    $h' \gets $ a hexahedron in $H$ containing $q'$ and not in $\textsc{Seen}$\;

    \uIf{there is such a hexahedron $h'$}{
      $(\sigma_{\text{old}}, \sigma^{-1}_{\text{old}}) \gets (\sigma, \sigma^{-1})$\;
      \ForEach{vertex $v$ of the quadrangle of $h$ opposite to $q$}{
        $v' \gets $ the corresponding vertex in $h'$\;
        \uIf{$\sigma(v)$ is undefined and $\sigma^{-1}(v')$ is undefined}{
          $\sigma(v) \gets v'$ \tcp*{Extend the map $\sigma$}
          $\sigma^{-1}(v') \gets v$\;
        }
        \ElseIf{$\sigma(v) \ne v'$ or $\sigma^{-1}(v') \ne v$}{
          $\textsc{Success} \gets \mathbf{false}$\;
          \textbf{break} \tcp*{Stop upon contradiction}
        }
      }

      \uIf{$\textsc{Success}$}{$\textsc{Seen} \gets \textsc{Seen} \cup \{h'\}$}
      \Else{$(\sigma, \sigma^{-1}) \gets (\sigma_{\text{old}},
        \sigma^{-1}_{\text{old}})$\;}
    }
    \Else{
      $\textsc{Success} \gets \mathbf{false}$\;
    }

    \uIf{the symbol after $h$ in $S$ is the branching symbol}{
      \tcc{All subsequent no-goods contain $h$. The search is aborted if its
        symmetric counterpart is not present.}
      \lIf{$\textsc{Success}$ is false}{\KwRet{\textbf{false}}}
    }
    \ElseIf{$\textsc{Success}$}{
      \KwRet{\textbf{true}} \tcp*{No-good is contained in $H$}
    }
  }
  \KwRet{\textbf{false}}\;
\end{algorithm}

\begin{algorithm}
  \caption{$\textsc{Try-Reverse-Shell}$: Determine whether or not a sequence of
    quad flips can create a given set of hexahedra}
  \label{alg:reverse-shell}
  \KwIn{%
    $Q$: a quadrangulation of the sphere; $H$: a set of hexahedra; $M$:
    maximum number of permutations to test.
  }
  \KwOut{%
    $\mathbf{true}$ if a sequence of quad flips was found.
  }

  \lIf{$H = \emptyset$}{\KwRet{\textbf{true}}}
 \ForEach{$h \in H$}{
    \If{$h$ can be added by performing a quad flip or $Q$ is a cube}{
      $Q' \gets $ the boundary after removing $h$\;
      \uIf{$\textsc{Try-Reverse-Shell}(Q', H \setminus \{h\}, M)$}{
        \KwRet{\textbf{true}}\;
      }
      \Else{
        Increment the number of tested permutations\;
      }
    }

    \If{$M$ permutations or more have been tested}{
      \KwRet{\textbf{false}}\;
    }
  }
  \KwRet{\textbf{false}}\;
\end{algorithm}

\section{Finding larger solutions using pre-computed meshes}
\label{sec:look-ahead}

The exhaustive search described in the previous section can only be used with
small limits on the maximum number of hexahedra, because of its exponential
execution time. In many cases, finding a complete shelling by searching
exhaustively is too difficult: the sequence of flips to construct the smallest
solution is too long, and the search tree contains many paths which transform
the initial boundary into one which is more difficult to mesh, instead of being
closer to a solution.

Instead of searching for a sequence of quad flips that transforms the initial
boundary $Q$ into a cube, the key idea for solving larger cases is to stop the
algorithm when a known configuration is found. For that purpose, we compute all
boundaries that can be shelled with at most $n$ hexahedra (say $n \le
11$). Using a list of all such boundaries and one of their shellings
(\autoref{sec:small-shellings}), this variant of the algorithm can efficiently
look up boundaries in the list during the search. This allows complete solutions
to be constructed from any sequence of flips leading to any of the boundaries in
the pre-computed set.

\subsection{Computing small shellable meshes}
\label{sec:small-shellings}

Consider the flip graph for quadrangulations of the sphere: its nodes represent
quadrangulations of the sphere, and arcs between these nodes represent a flip
between two quadrangulations. A breadth-first traversal of this graph starting
from the cube and stopped at depth $n$ generates all quadrangulations that can
be obtained using a sequence of up to $n$ flips. To deal with cycles in this
graph, previously visited quadrangulations are stored in a hash table. The hash
value for quadrangulations is constructed from a signature based on the valence
of vertices, and the isomorphism test two quadrangulations is performed using a
variation on \autoref{alg:symmetry} where the two starting quadrangles are part
of different quadrangulations.

\begin{algorithm}
  \caption{$\textsc{Generate-Shellings}$: Generate small shellable hexahedral meshes}
  \label{alg:generate-small}

  \KwIn{$n$: maximum size for the generated hexahedral meshes}
  \KwOut{$\mathcal{H}$: a set of hexahedral shellings with up to $n$ hexahedra
    in each mesh.}

  $S \gets \emptyset$\;
  $\mathcal{H} \gets \emptyset$\;
  $Q \gets \textsc{New-Queue}()$\;
  $\textsc{Enqueue}(Q, \textsc{Cube})$\;
  \While{$Q$ is not empty}{
    $H \gets \textsc{Dequeue}(Q)$\;
    $\mathcal{H} \gets \mathcal{H} \cup \{H\}$\;
    \lIf{$|H| = n$}{\textbf{continue}}
    \ForEach{quad flip $F$}{
      $(B, h) \gets \textsc{Perform-Flip}\left(F, \partial{}H\right)$\;
      \If{$\textsc{Is-Compatible}(H, h) \land B \not\in S$}{
        $S \gets S \cup \{B\}$\;
        $\textsc{Enqueue}(Q, H \cup \{h\})$\;
      }
  }
  }
  \KwRet{$\mathcal{H}$}\;
\end{algorithm}

The signature used by our algorithm is a histogram of the valences of each
vertex, followed by the number of edges connecting vertices of valence $v_a$ and
valence $v_b$, for any $v_a$ and $v_b$ where this number is non-zero. While this
choice of signature causes collisions, it can be computed quickly and new
entries can be inserted without necessarily needing a slower computation to find
a unique canonical representation.

Not all quadrangulations generated in this breadth first search admit a shelling
with up to $n$ hexahedra: interpreting the flips performed during the traversal
of the graph as the insertion of hexahedra, these hexahedra may not all be
compatible. By explicitly testing for compatibility while performing the breadth
first search (\autoref{alg:generate-small}), we obtain a greedy construction
similar to the procedure outlined by \citet{Xiang-2018}: the hexahedra that are
found are those which admit a shelling such that any prefix is the smallest
shellable hexahedral mesh for the corresponding boundary. For large values of
$n$, it is not clear that such a shelling should always exist, but we can verify
this property for small values of $n$. For every quadrangulation found during
the breadth-first search but without a hexahedral mesh found by
\autoref{alg:generate-small}, \autoref{alg:search-exhaustive} is used to verify
that there is indeed no shellable hexahedral mesh with at most $n$
hexahedra. This test was performed for $n \le 10$, and no counter-examples were
found.

From \autoref{alg:generate-small}, a table of $69,043,690$ boundaries that can
be meshed with up to 11 hexahedra is constructed (\autoref{table:quadrangulations}).

\begin{table}
  \centering
  \caption{Number of combinatorial quadrangulated boundaries that can be shelled 
    with up to $H_{\max}$ hexahedra.
    Timings are given for a single thread on an
    Intel\textregistered{} Core\texttrademark{} i7-7700HQ CPU.}
  \label{table:quadrangulations}
  \begin{tabular}{rrl}
     $H_{\max}$ & \# quad meshes & timing \\
    \hline
    1  & 1          & $< 0.1$s      \\
    2  & 2          & $< 0.1$s      \\
    3  & 5          & $< 0.1$s      \\
    4  & 17         & $< 0.1$s      \\
    5  & 74         & $< 0.1$s      \\
    6  & 489        & $< 0.1$s      \\
    7  & 4,192      & $0.12$s       \\
    8  & 42,676     & $1.78$s       \\
    9  & 476,520    & $34.418$s     \\
    10 & 5,632,488  & $14$min $55$s \\
    11 & 69,043,690 & $6$h $41$min  \\
  \end{tabular}
\end{table}

\subsection{Using the pre-computed table}
\label{sec:use-list}

If at any point during the search, the boundary of the unmeshed region
matches one of the pre-computed quadrangulations, the shelling of that
quadrangulation is used to finish the meshing of that region.

\begin{figure}
  \includegraphics[width=0.8\linewidth]{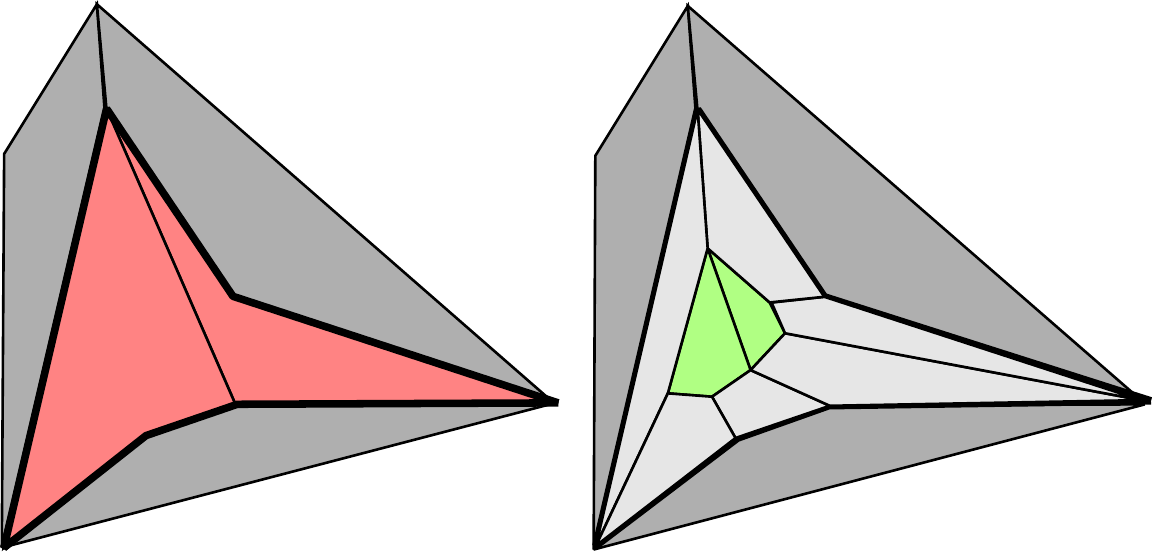}
  \caption{Principle of the insertion of a buffer layer to guarantee final mesh
    validity when using precomputed cavity meshes.  (left) Simply adding two
    quadrangles (red) inside a cavity creates an invalid mesh where pairs of
    quads share more than one facet; (right) inserting a layer of buffer quads
    (light gray) allows the use of the same (combinatorial) quads to fill the
    cavity and produce a valid mesh.}
  \label{fig:buffer-cavity}
\end{figure}

The idea is to use the shelling computed in the previous section to fill the
unmeshed region. Simply combining the two solutions is not always possible: this
may produce an invalid mesh where, for example, two hexahedra share multiple
quadrangles (\autoref{fig:buffer-cavity}). \autoref{alg:search-exhaustive} could
be used to compute all shellings of the unmeshed region with up to $n$
hexahedra. If this search finds a shelling compatible with the partial solution
constructed so far, a solution can be generated, at the cost of an additional
computation.

Even if no such shelling was found, a solution can be constructed from any
shelling of the unmeshed region, without performing an additional search or
storing multiple hexahedrizations for each boundary: first construct a copy of
the boundary of the unmeshed region, then, for each quadrangle of this boundary,
create a hexahedron to connect each quadrangle to its copy. The hexahedra that
have been inserted in this manner are guaranteed to be compatible with any
hexahedrization of the unmeshed region, allowing a complete mesh to be
constructed.  When this approach is used, the first solution found by the
algorithm is not in general the smallest. However, when the smallest solution
contains a large number of hexahedra, this approach can construct solutions in
many cases where methods with stronger guarantees fail to find any, because it
adds several hexahedra without branching.

Efficient access to the pre-computed table is performed using a binary
search. We create an array of all the quadrangulations we found, sorted by their
signatures. To find the hexahedral mesh corresponding to a given
quadrangulation, its signature is computed and an isomorphism test is performed
on all quadrangulations in the table that have the same signature.

\section{Results}

\paragraph{A constructive solution for constrained hex-meshing}

Only one previous solution to the constrained hexahedral meshing problem gives a
completely explicit construction \cite{Carbonera-2010}. This method requires
$5396~n$ hexahedra to construct a valid mesh bounded by $n$ quadrangles. In the
following, we prove that this bound can be lowered to $78~n$ using the
construction proposed by \citet{Erickson-2014}.

\begin{figure*}[h]
  \centering
  \includegraphics[width=0.2\linewidth]{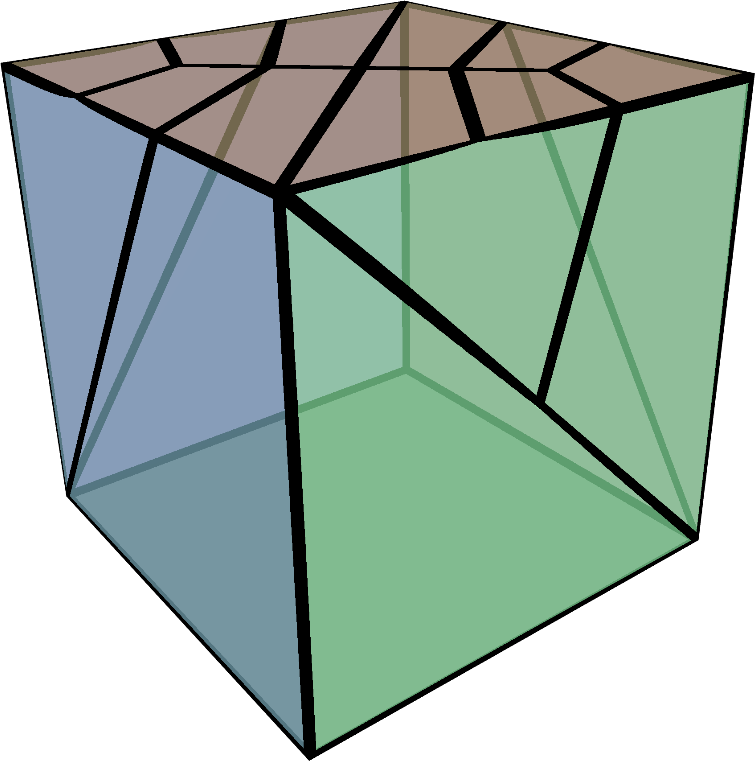}
  \includegraphics[width=0.2\linewidth]{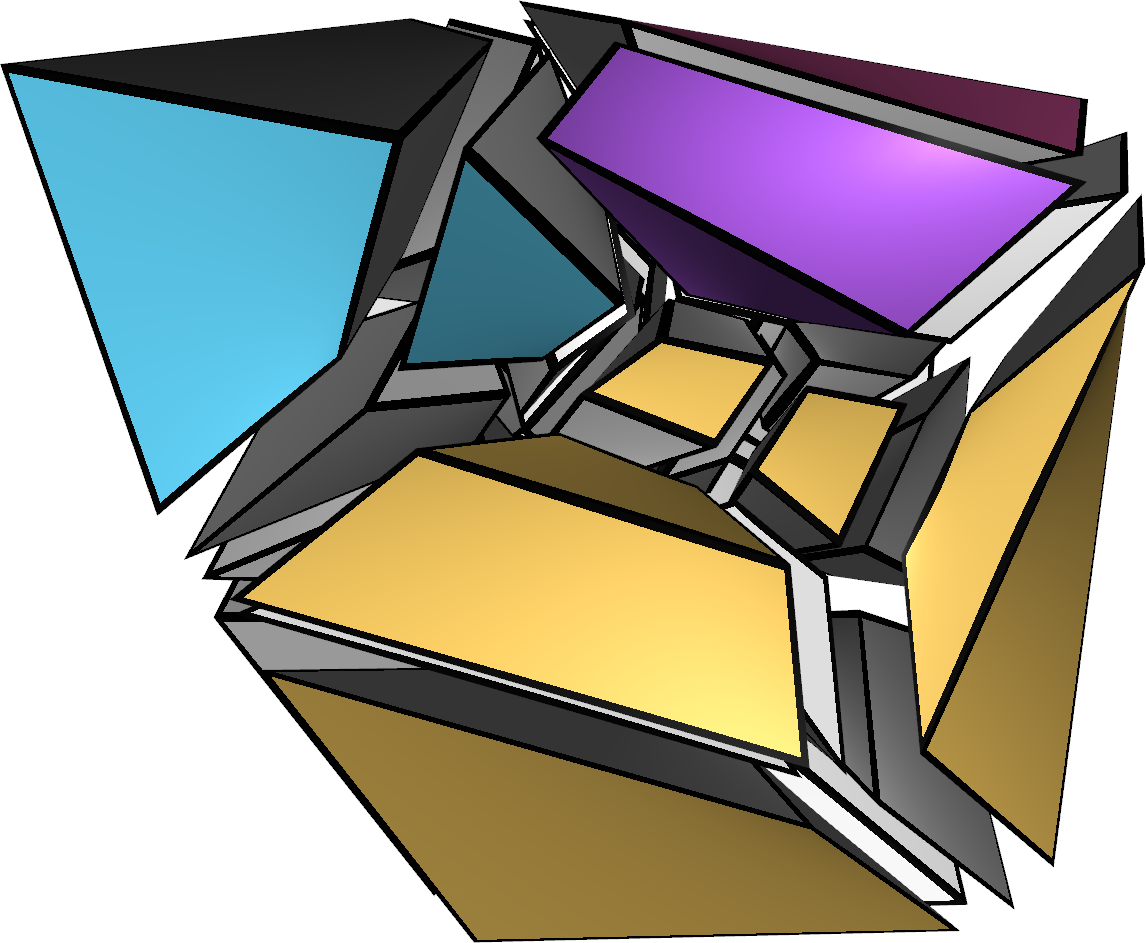}
  \includegraphics[width=0.15\linewidth]{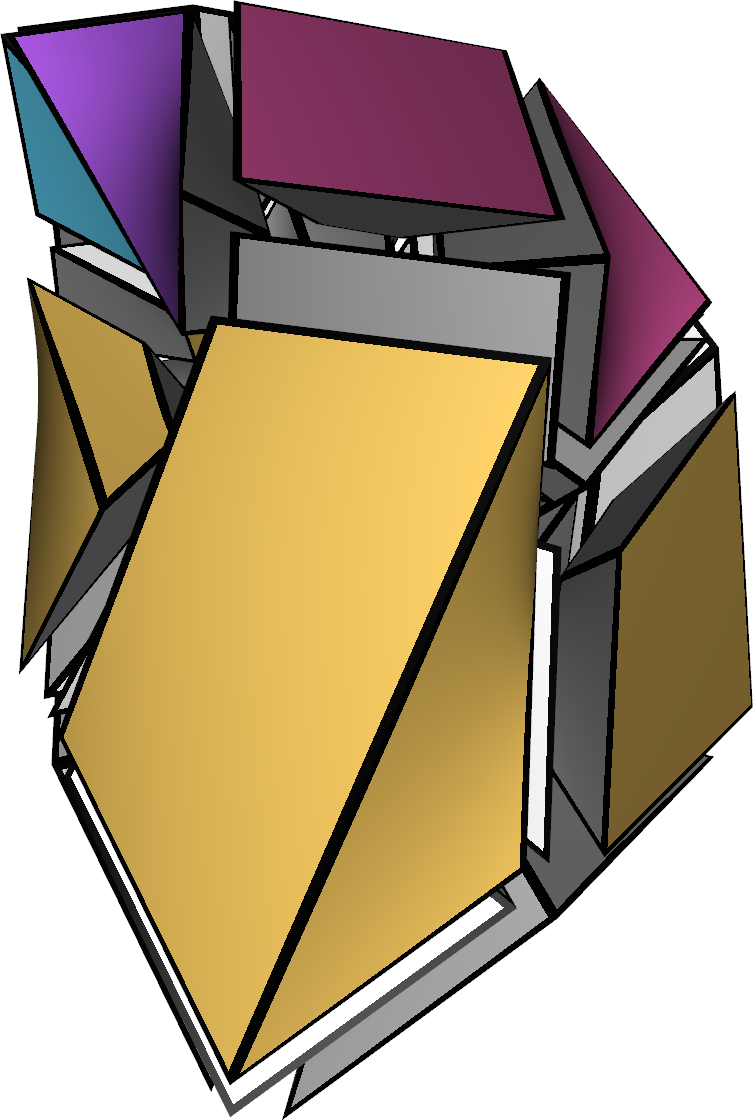}
  \includegraphics[width=0.2\linewidth]{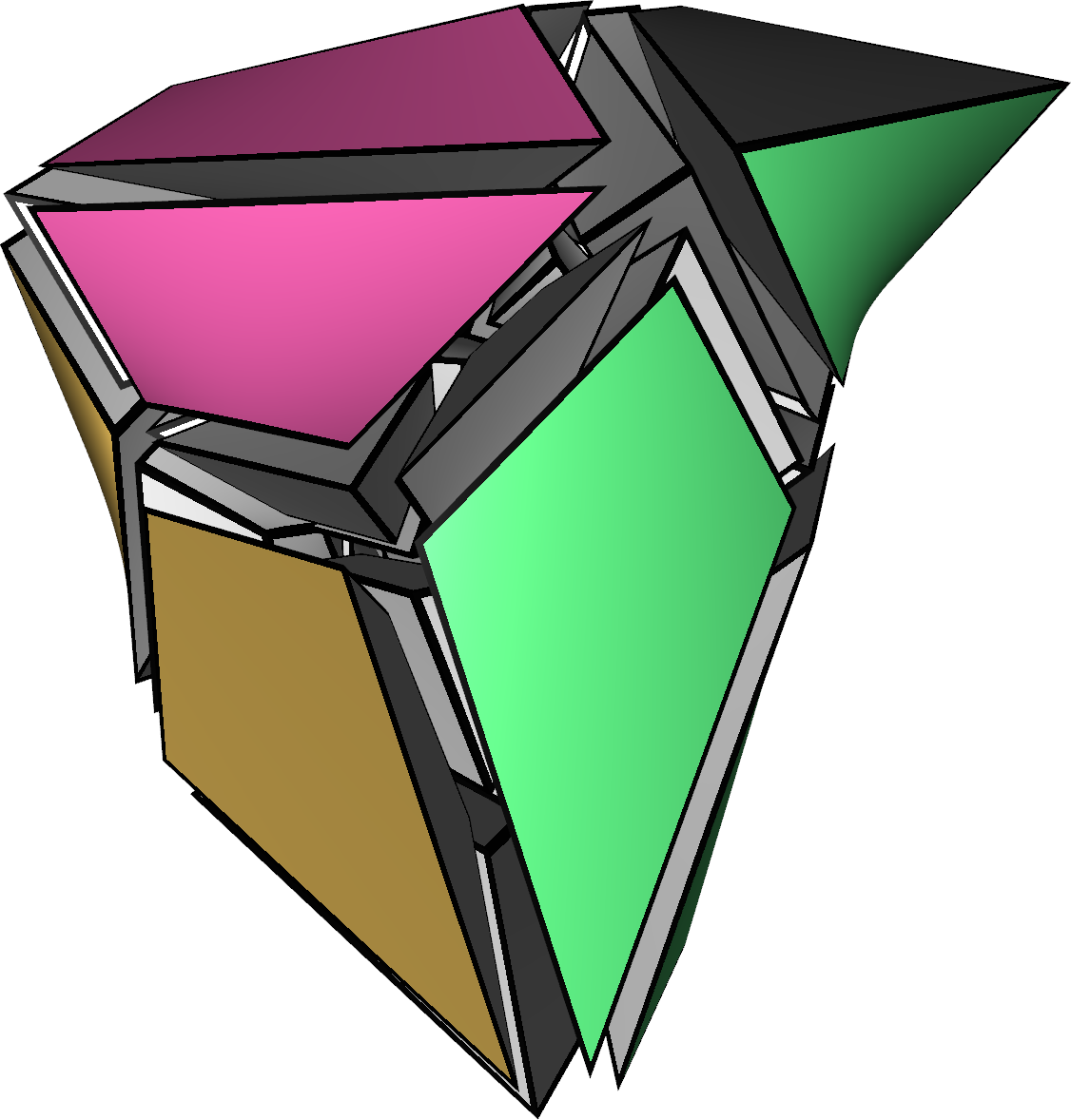}
  \includegraphics[width=0.2\linewidth]{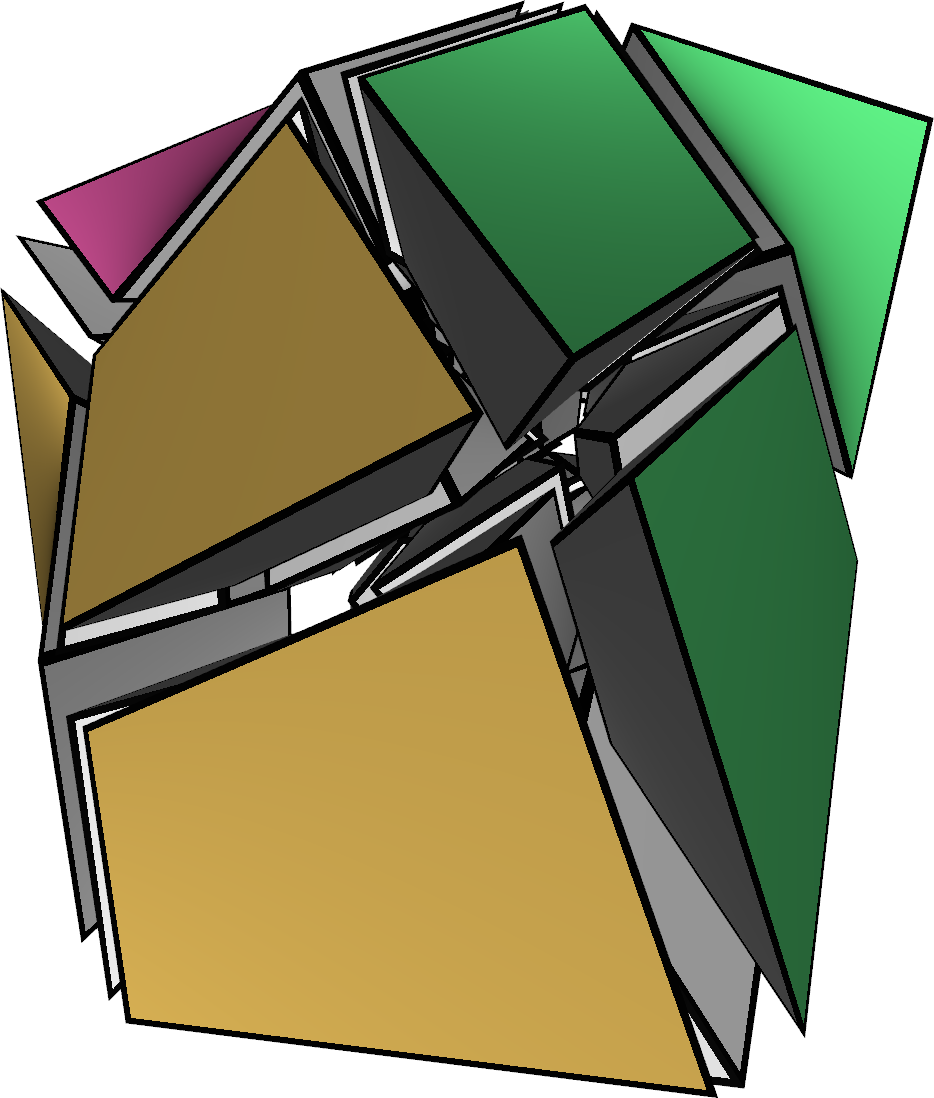}%
  \vspace{.7cm}

  \includegraphics[width=0.2\linewidth]{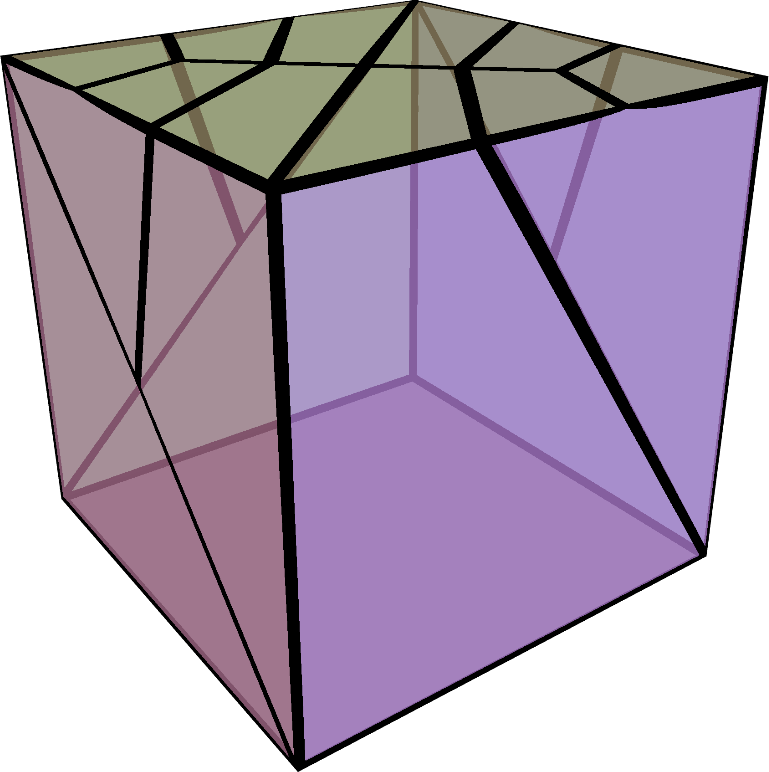}
  \includegraphics[width=0.2\linewidth]{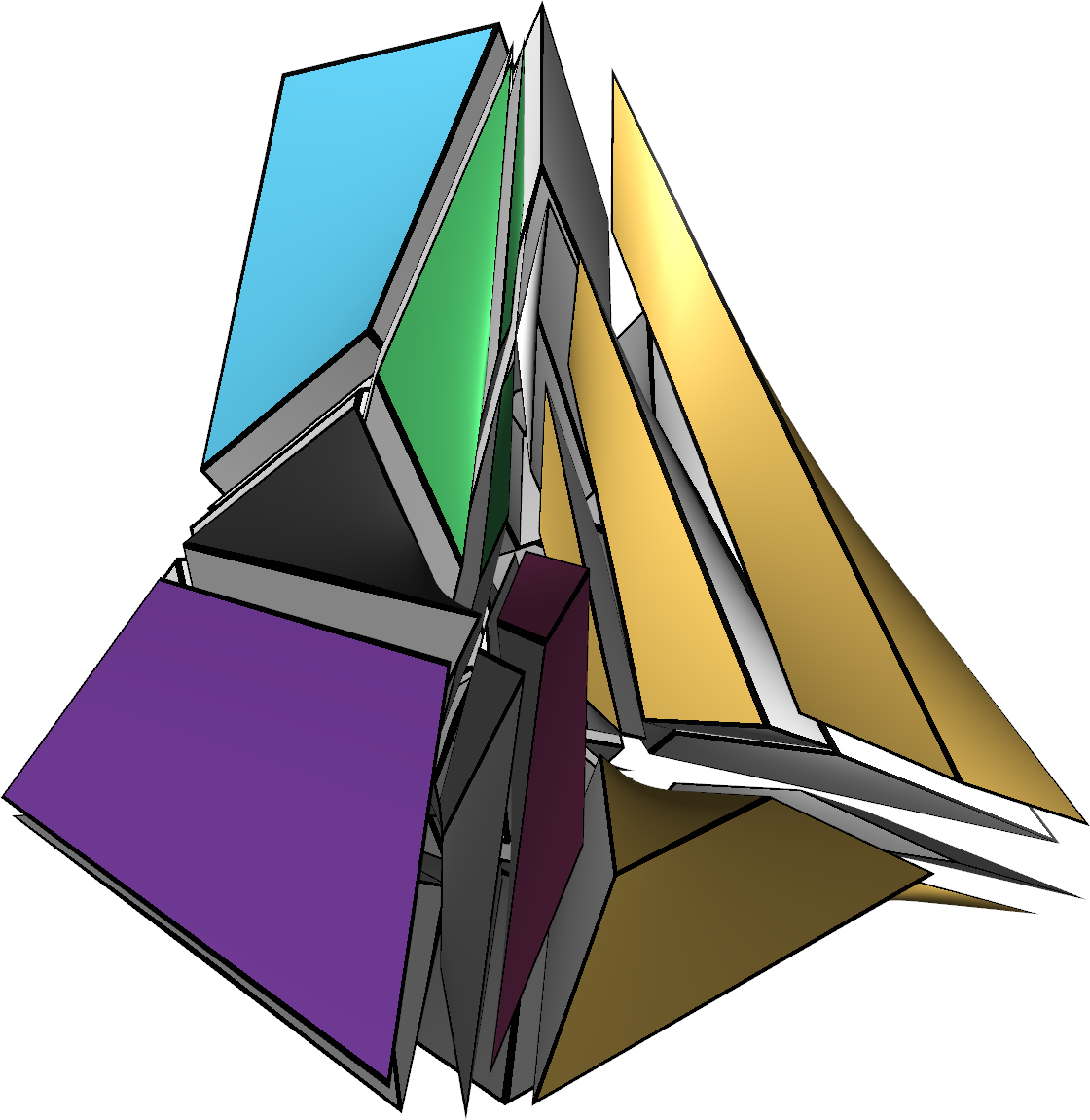}
  \includegraphics[width=0.2\linewidth]{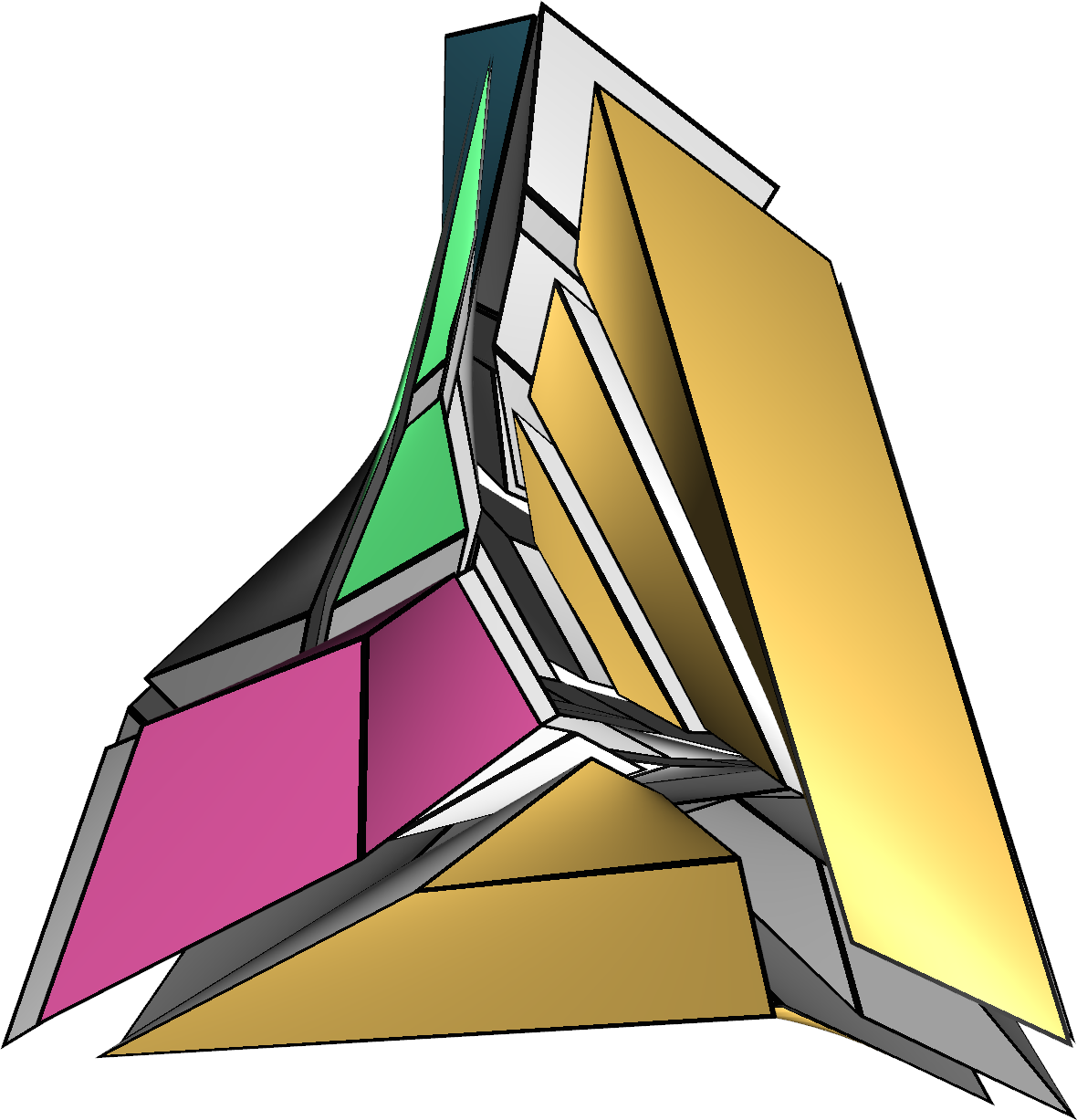}
  \includegraphics[width=0.15\linewidth]{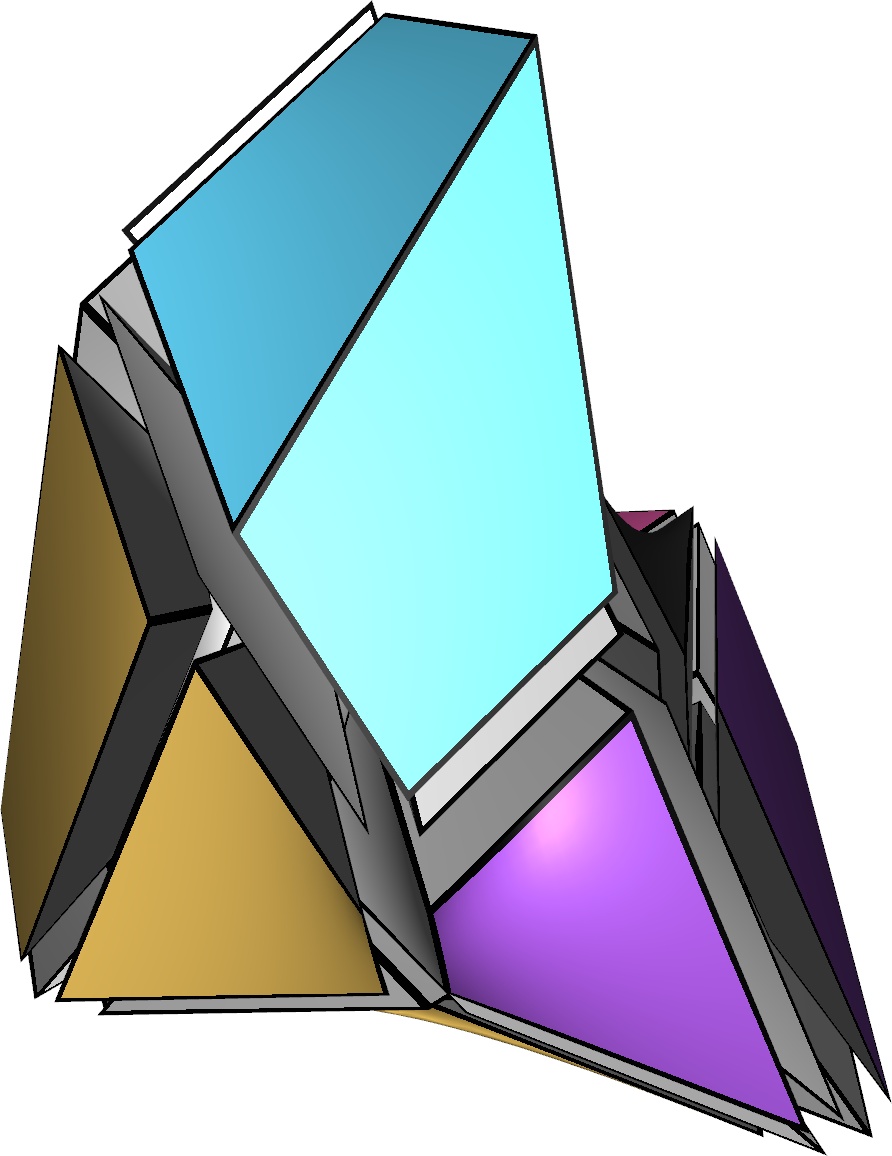}
  \includegraphics[width=0.2\linewidth]{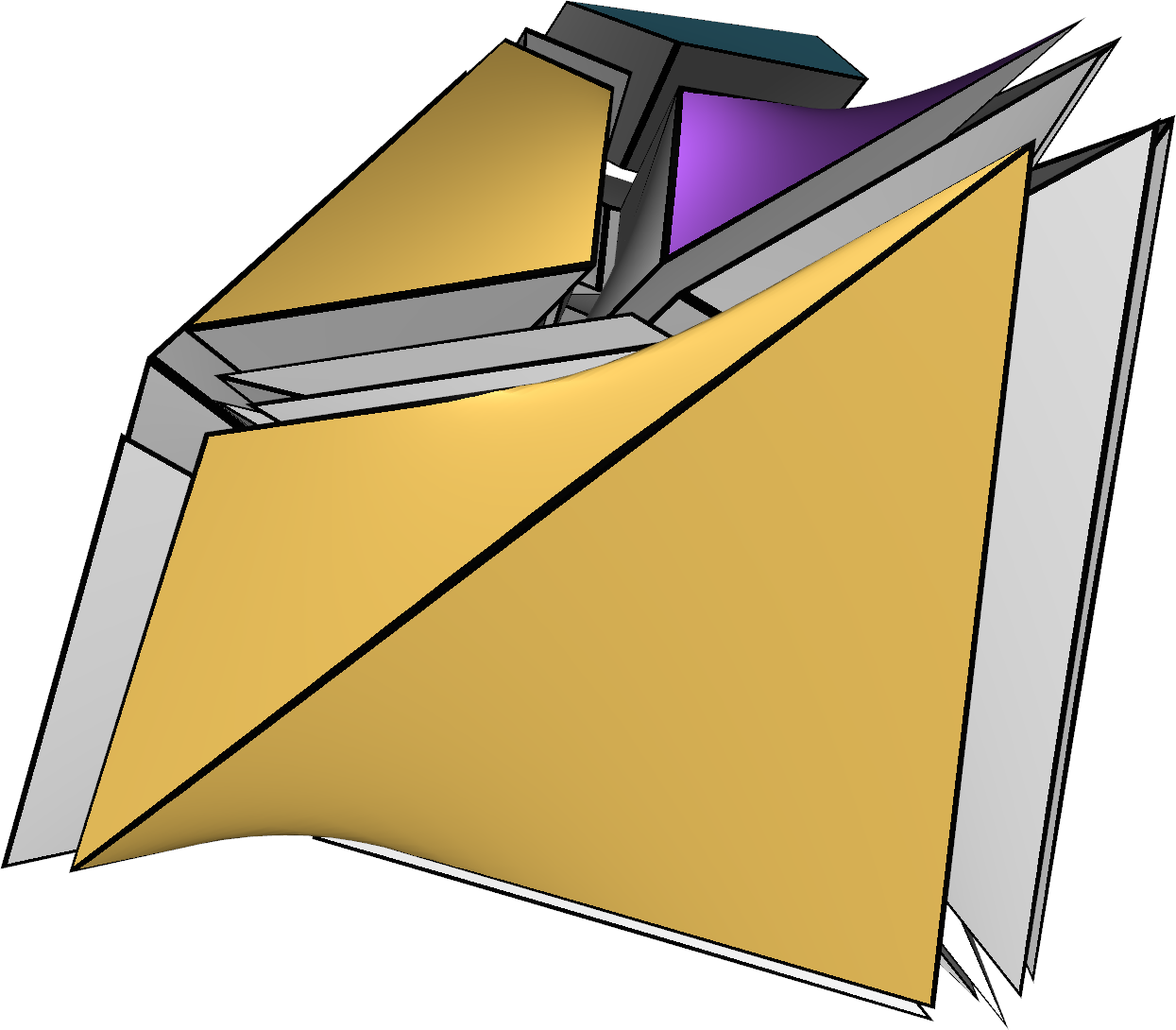}
  \vspace{.2cm}

  \caption{Hexahedrizations of the two types of buffer cubes used to mesh
    arbitrary domains in the algorithm of \citet{Erickson-2014}. (top) 37
    hexahedra to mesh the 20-quadrangle cell; (bottom) 40 hexahedra to mesh the
    22-quadrangle cell. Colors correspond to the different sides of the
    original cubes (shown on the left).}
  \label{fig:erickson-buffers-hex}
\end{figure*}

Using our search algorithm (\autoref{sec:look-ahead}), we found hexahedral
meshes for both types of buffer cells that Erickson's construction needs, along
with geometric realizations using linear hexahedra
(\autoref{fig:erickson-buffers-hex}) obtained by applying existing mesh
untangling techniques \cite{Toulorge-2013, Livesu-2015}, although the solutions
have a very low minimum scaled Jacobian (\autoref{table:geometric-stats}). The
meshes that we found contain 37 and 40 hexahedra. Because gluing multiple buffer
cells together as needed by the construction would create a degenerate mesh, a
hexahedron is added on each boundary quadrangle. The resulting meshes of the
buffer cells have 57 and 62 hexahedra respectively, giving the following result:

\begin{theorem}
  Let $\Omega$ be a compact and connected subset of $\mathbb{R}^3$ bounded by a
  2-manifold $\partial\Omega$. Given a quadrangulation $Q$ of $\partial\Omega$,
  each component of $Q$ containing an even number of quadrangles, and a
  triangulation $T$ of $\Omega$ (splitting each quadrangle of $Q$ into two
  triangles), if there is a combinatorial hexahedral mesh of $\Omega$ bounded by
  $Q$, then there is one with no more than $62|Q| + 8|T|$ hexahedra. In
  particular, if $\Omega$ is a ball (hence $\partial\Omega$ is a sphere) and
  $|Q|$ is even, there is a combinatorial hexahedral mesh bounded by $Q$ with no
  more than $78|Q|$ hexahedra.
  \label{theorem:hex-bound}
\end{theorem}

\begin{proof}
  Follow the construction of \citet{Erickson-2014} using the templates that
  we computed. There is one buffer cell for each boundary quadrangle, and each
  tetrahedron of the triangulation $T$ is split into 4, 7, or 8 hexahedra. In
  the worst case, each buffer cell will be meshed with $62$ hexahedra, and each
  tetrahedron will be split into $8$ hexahedra.

  If $\Omega$ is a ball, there is always a triangulation $T$ with $2|Q|$
  tetrahedra, obtained by arbitrarily splitting each quadrangle into two
  triangles, adding a vertex inside the domain, and joining each triangle to
  this new vertex by a tetrahedron. The bound for this special case is therefore
  $62|Q| + 8 \times 2|Q| = 78|Q|$.
\end{proof}

A similar bound can be obtained for quadrangulations with an odd-number of
quadrangles in some of their components. In that case, hexahedra are added to
connect pairs of odd components, and \ref{theorem:hex-bound} is used to compute
the number of hexahedra to mesh the rest of the domain.

\begin{table*}
  \caption{Statistics for the geometric meshes computed for the test cases
  of Figures~\ref{fig:pyramid-spindle}~and~\ref{fig:erickson-buffers}}
  \label{table:geometric-stats}
  \begin{tabular}{l cccc ccc ccc}
    \multirow{2}{*}{Template} & \multirow{2}{*}{$|Q|$} &
                                                          \multirow{2}{*}{$|V_{\text{bnd}}|$}
    & \multirow{2}{*}{$|H|$} & \multirow{2}{*}{$|V_{\text{total}}|$} &
                                                                       \multicolumn{3}{c}{\# edges per valence} & \multicolumn{3}{c}{Scaled Jacobian} \\
                               & & &  &  & 3 & 4 & 5 & min & max & median \\
    \hline
    Tetragonal trapezohedron   & 8  & 10 & 40 & 52 & 40 & 75 & 4 & 0.35 & 0.42 & 0.38 \\
    Schneiders' pyramid        & 16 & 18 & 36 & 51 & 32 & 62 & 4 & 0.12 & 0.49 & 0.26 \\
    Erickson's buffer cell (1) & 20 & 22 & 37 & 53 & 43 & 49 & 4 & 0.31 & 0.63 & 0.42 \\
    Erickson's buffer cell (2) & 22 & 24 & 40 & 55 & 44 & 48 & 9 & 0.031 & 0.45 & 0.41 \\
  \end{tabular}
\end{table*}

\paragraph{Hexahedrizations for small quadrangulations of the sphere}

\begin{figure}
  \centering
  \vspace{.5cm}
  \includegraphics[width=\linewidth]{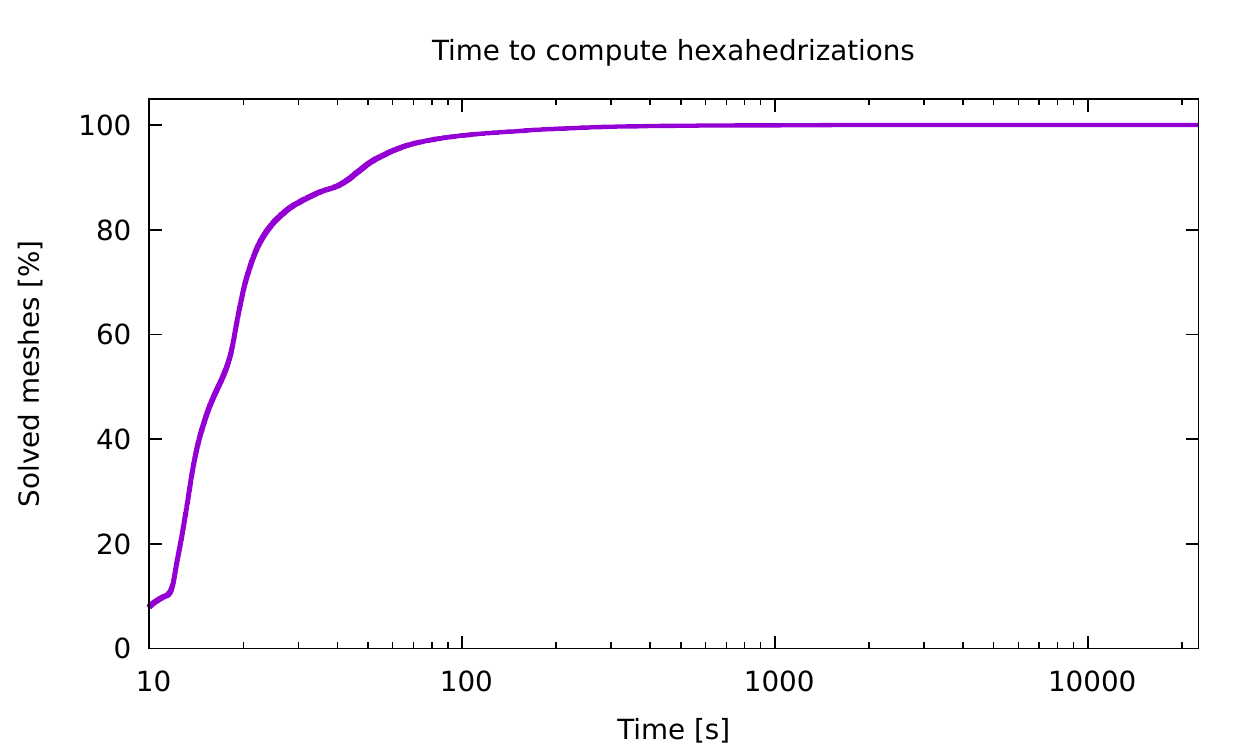}
  \caption{Time to compute hexahedrizations for all quadrangulations of the
    sphere with up to 20 quadrangles. Run on a machine with two AMD EPYC 7551
    CPUs (32 cores per CPU).}
  \label{fig:time-quadrangulations}
\end{figure}

\begin{figure}
  \centering
  \vspace{.5cm}
  \includegraphics[width=\linewidth]{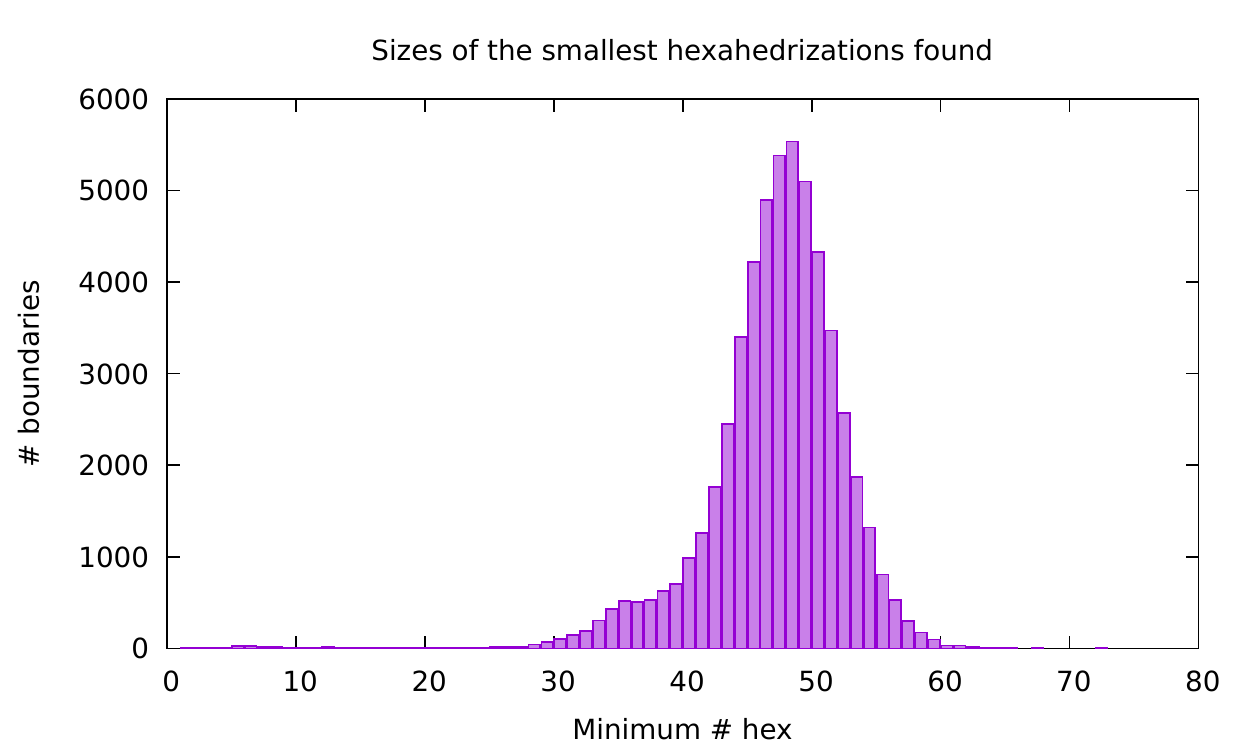}
  \caption{Sizes of the smallest hexahedrizations found for all quadrangulations of the
    sphere with up to 20 quadrangles.}
  \label{fig:size-quadrangulations}
\end{figure}

\begin{figure}
  \centering
  \includegraphics[width=0.39\linewidth]{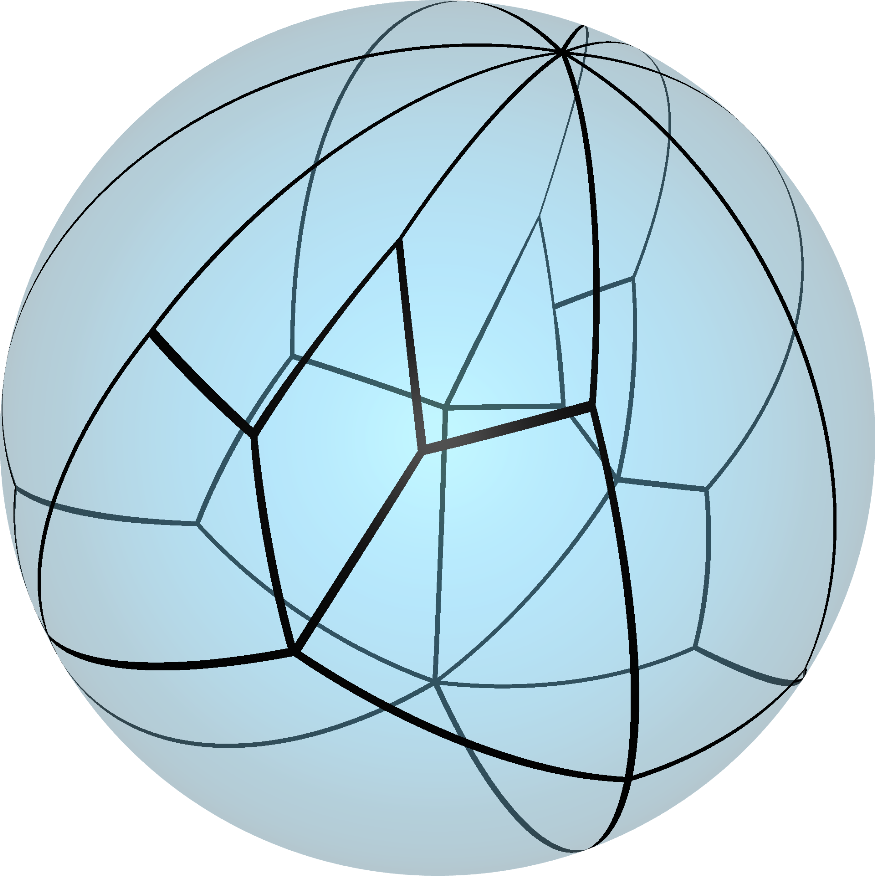}~%
  \hspace{.5cm}
  \includegraphics[width=0.39\linewidth]{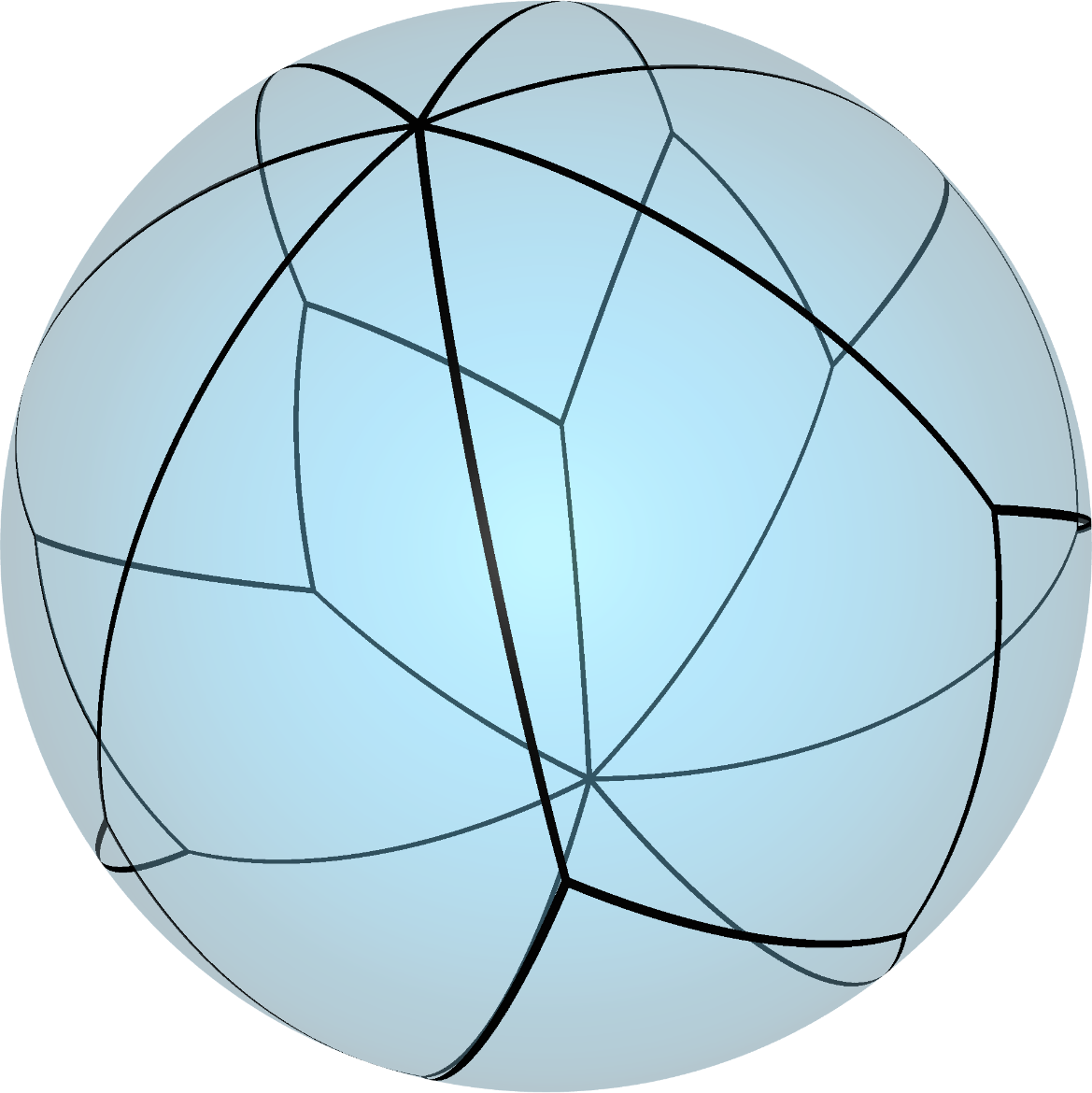}%
  \vspace{.5cm}
  \includegraphics[width=0.39\linewidth]{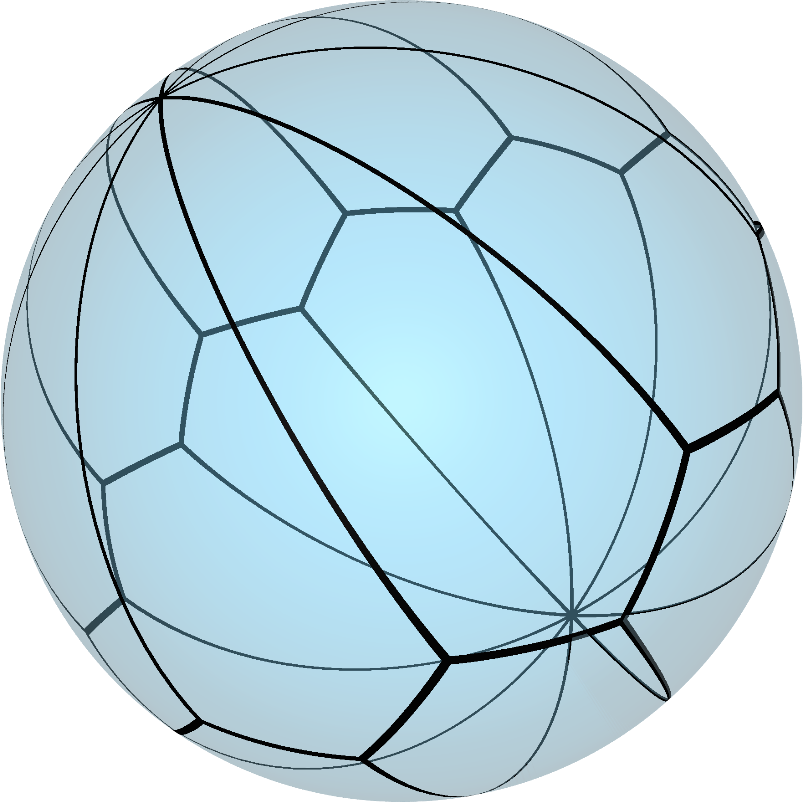}~%
  \hspace{.5cm}
  \includegraphics[width=0.39\linewidth]{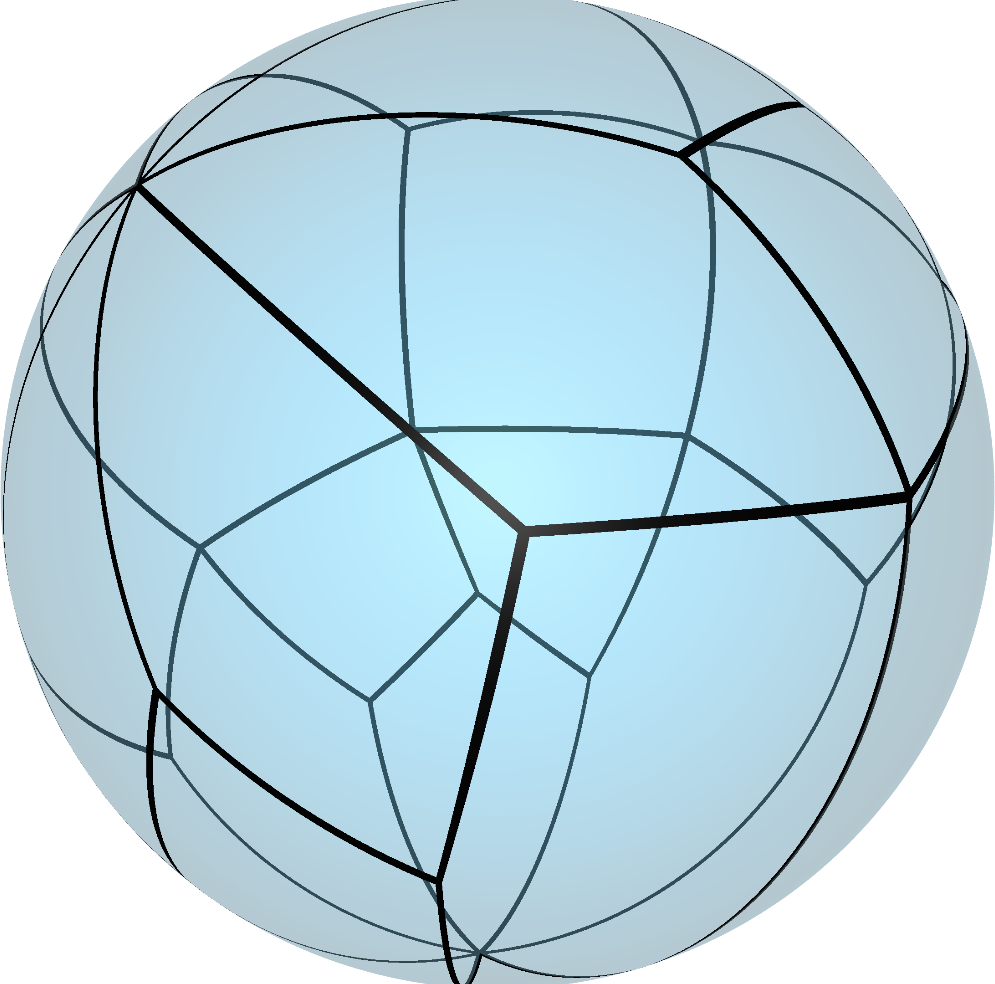}%
  \caption{The four most time-consuming quadrangulated spheres to mesh using our
    method. Each required over an hour of computation time on 64 cores.}
  \label{fig:pathological-cases}
\end{figure}

We used the algorithm described in \autoref{sec:look-ahead} to compute
hexahedrizations for all even quadrangulations of the sphere containing up to
$20$ quadrangles (\autoref{table:quad-20-stats}). The $54,943$ input
quadrangulations were generated using \texttt{plantri} \cite{Brinkmann-2005}. We
pre-computed shellable hexahedral meshes with up to 11 hexahedra.  Of the
$69,043,690$ boundaries that were pre-computed, only $130$ are included in the
list of inputs. Nonetheless, in about 20\% of all instances, the search for a
solution terminates almost immediately after loading the set of pre-computed
solutions (\autoref{fig:time-quadrangulations}). Only a few additional seconds
are enough to find hexahedrizations bounded by most quadrangulations of the
sphere. There are however some more difficult cases, requiring over an hour of
computation time (\autoref{fig:pathological-cases}). The trapezohedron bounded
by $n$ faces, obtained by generalizing the tetragonal trapezohedron of
\autoref{fig:pyramid-spindle}, is usually among the most difficult cases of a
given size, requiring meshes with an intricate internal structure in order to be
filled. For example, the smallest solution found for the 20-face decagonal
trapezohedron contained $72$ hexahedra, strictly more than any of the other
boundaries (\autoref{fig:size-quadrangulations}). Similarly, the 16-face
octagonal trapezohedron required $67$ hexahedra, with the decagonal
trapezohedron being the only boundary for which all solutions found were
larger. The trapezohedra are also among the boundaries that require the most
time before any solution could be found. The 14-face heptagonal trapezohedron is
the second most time consuming input, requiring 2h 50min, and the 20-face
decagonal trapezohedron is the third, requiring 2h 43min. In the worst case,
shown on \autoref{fig:pathological-cases}, it took 6h 15min before a 58-element
mesh was found.

The quadrangulations of \autoref{fig:pyramid-spindle} are not particularly
difficult to mesh using our method. Indeed, most quadrangulations of up to 18
quadrangles require more computation time than those two cases --- and even
finding the smallest known solutions is orders of magnitude easier than finding
any solutions for the cases shown on \autoref{fig:pathological-cases}. On a
4-core Intel\textregistered{} Core\texttrademark{} i7-7700HQ CPU, after
pre-computing a list of shellable meshes with up to 10 hexahedra, the 36-element
mesh of Schneiders' pyramid originally found by \citet{Xiang-2018} is found
within 3 seconds of search. A 44-element mesh of the tetragonal trapezohedron is
found within 4 seconds and it takes 31 seconds to find the smallest known
40-element mesh constructed in \cite{Verhetsel-2018}. Statistics for the
geometric realizations found for both meshes are shown on
\autoref{table:geometric-stats}.

\begin{table}
  \centering
  \caption{Statistics for the combinatorial meshes computed for all
    even quadrangulations of the sphere with up to 20 quadrangles.}
  \label{table:quad-20-stats}
  \begin{tabular}{c ccc ccc cccc}
    \multirow{2}{*}{$Q$} &
                                                                   \multicolumn{3}{c}{$|H|$}
    & \multicolumn{3}{c}{$|V_{\text{total}}|$} & \multicolumn{4}{c}{\%edges by valence} \\
                           & min & max & med & min & max & med & 3 & 4 & 5 & 6 \\
    \hline
    6 & \multicolumn{3}{c}{$1$} & \multicolumn{3}{c}{$8$} & \multicolumn{4}{c}{(none)} \\
    8 & \multicolumn{3}{c}{$44$} & \multicolumn{3}{c}{$56$} & 48 & 34 & 16 & 2 \\
    10 & 2 & 58 & 36 & 12 & 64 & 48 & 39 & 45 & 15 & 1 \\
    12 & 3 & 47 & 43 & 14 & 57 & 52 & 38 & 50 & 11 & 1 \\
    14 & 3 & 59 & 44 & 16 & 73 & 55 & 38 & 48 & 12 & 1 \\
    16 & 4 & 67 & 45 & 18 & 77 & 56 & 37 & 51 & 11 & 1 \\
    18 & 4 & 67 & 46 & 20 & 79 & 58 & 38 & 49 & 12 & 1 \\
    20 & 5 & 72 & 47 & 22 & 81 & 59 & 38 & 49 & 12 & 1 \\
  \end{tabular}
\end{table}

% \begin{table*}
%   \centering
%   \caption{Statistics for the combinatorial meshes computed for all
%     even quadrangulations of the sphere with up to 20 quadrangles.}
%   \label{table:quad-20-stats}
%   \begin{tabular}{lllllllllllll}
%     \multirow{2}{*}{$|Q|$} & \multirow{2}{*}{$|V_{\text{bnd}}|$} &
%                                                                    \multicolumn{3}{c}{$|H|$}
%     & \multicolumn{3}{c}{$|V_{\text{total}}|$} & \multicolumn{4}{c}{\# edges
%                                                  grouped by valence} \\
%                            & & min & max & median & min & max & median & 3 & 4 & 5 & 6 \\
%     \hline
%     6 & 8 & \multicolumn{3}{c}{$1$} & \multicolumn{3}{c}{$8$} & \multicolumn{4}{c}{(no interior edges)} \\
%     8 & 10 & \multicolumn{3}{c}{$44$} & \multicolumn{3}{c}{$56$} & 48\% & 34\% & 16\% & 2\% \\
%     10 & 12 & 2 & 58 & 36 & 12 & 64 & 48 & 39\% & 45\% & 15\% & 1\% \\
%     12 & 14 & 3 & 47 & 43 & 14 & 57 & 52 & 38\% & 50\% & 11\% & 1\% \\
%     14 & 16 & 3 & 59 & 44 & 16 & 73 & 55 & 38\% & 48\% & 12\% & 1\% \\
%     16 & 18 & 4 & 67 & 45 & 18 & 77 & 56 & 37\% & 51\% & 11\% & 1\% \\
%     18 & 20 & 4 & 67 & 46 & 20 & 79 & 58 & 38\% & 49\% & 12\% & 1\% \\
%     20 & 22 & 5 & 72 & 47 & 22 & 81 & 59 & 38\% & 49\% & 12\% & 1\% \\
%   \end{tabular}
% \end{table*}

\section{Conclusions}

While existence proofs for solutions to the constrained hexahedral meshing
problem have long existed, previously known methods to construct hexahedral
meshes have required very large meshes even for small quadrangulations; this
paper shows that a much lower theoretical bound exists, and provides tools to
search for much smaller hexahedral meshes.

Cases previously thought of as difficult and which motivated research about this
question are solved in a matter of seconds using our techniques. This research,
by allowing hexahedral meshes to be computed for any small quadrangulation of
the sphere, opens up a wide array of new possibilities. It is an important step
for hex-dominant meshing \cite{Yamakawa-2003}, as this solves the combinatorial
aspect of the problem of filling the cavities that those methods leave. This
method also offers new insights into the structure of block decompositions for
configurations for which state-of-the-art techniques such as \cite{Liu-2018}
fail to generate a valid block structure, by allowing combinatorial meshes to be
computed in some of those cases.

\begin{acks}
  This research is supported by the \grantsponsor{ERC}{European Research
    Council}{} (project HEXTREME,
  \grantnum{ERC}{ERC-2015-AdG-694020}). Computational resources have been
  provided by the supercomputing facilities of the Université catholique de
  Louvain (CISM/UCL) and the Consortium des Équipements de Calcul Intensif en
  Fédération Wallonie Bruxelles (CÉCI) funded by the \grantsponsor{CECI}{Fond de
    la Recherche Scientifique de Belgique (F.R.S.-FNRS)}{} under convention
  \grantnum{CECI}{2.5020.11}.
\end{acks}

\bibliographystyle{ACM-Reference-Format}
\interlinepenalty=10000
\bibliography{paper}

\end{document}